\documentclass[aip,jcp,amsmath,amssymb,reprint]{revtex4-2}

\usepackage{graphicx}       
\usepackage{dcolumn}        
\usepackage{bm}             
\usepackage{hyperref}       
\usepackage{xcolor}         
\usepackage{microtype}      
\usepackage{upgreek}        
\usepackage{booktabs}       
\usepackage[obeyFinal]{todonotes} 
\usepackage{comment}

\definecolor{revblue}{RGB}{0,90,180}

\begin{document}

\title{Strong-Field-Driven Non-Linear Electron Dynamics in Thiophene Oligomers}

\author{Mustapha Driouech}
\email{m.driouech@uni-jena.de}
\affiliation{Friedrich-Schiller Universit\"at Jena, Institute for Condensed Matter Theory and Optics, 07743 Jena, Germany}
\author{Caterina Cocchi}
\email{caterina.cocchi@uni-jena.de}
\affiliation{Friedrich-Schiller Universit\"at Jena, Institute for 
Condensed Matter Theory and Optics, 07743 Jena, Germany}
\affiliation{Friedrich-Schiller Universit\"at Jena, Abbe Center of Photonics, Institute of Physics, 07745 Jena, Germany}
\author{Michele Guerrini}
\email{michele.guerrini@uni-jena.de}
\affiliation{Friedrich-Schiller Universit\"at Jena, Institute for Condensed Matter Theory and Optics, 07743 Jena, Germany}

\begin{abstract}
The interaction between conjugated molecules and intense electric field pulses drives a plethora of intriguing nonlinear phenomena, including optical limiting. While excited-state absorption was recently identified as the primary mechanism for this effect in thiophene oligomers, open questions remain regarding field-dependent population dynamics under intense laser driving. Using real-time time-dependent density functional theory combined with a determinant-overlap population framework, we track the intensity-dependent nonlinear response of a single thiophene ring (1T) and quaterthiophene (4T) as prototypical members of the oligothiophene family with different length and symmetry. We demonstrate that extended $\pi$-conjugation in 4T lowers nonlinear excitation thresholds by orders of magnitude compared to 1T. State-resolved population dynamics reveal complete ground-state depletion accompanied by sequential excited-state absorption channels under intense driving. High-harmonic generation spectra explicitly reflect molecular symmetry constraints, while energy-resolved occupation densities confirm continuous orbital redistribution across the $\pi^*$-manifold. This work provides dynamic physical insights for engineering organic materials with tailored nonlinear optical properties.
\end{abstract}

\maketitle
\section{Introduction}

Conjugated molecules stimulated by external time-dependent fields display a variety of linear and nonlinear excitations~\cite{dela+00cr}, making them promising candidates for applications in ultrafast photonics~\cite{flic+18nanophotonics}, optical limiting~\cite{tutt-bogg93pqe,sun-rigg99irpc,span99jmc}, and strong-field spectroscopy~\cite{gree+90sci,desi+18epjb}. Among these systems, oligothiophenes play a central role as fundamental building blocks of organic semiconductors~\cite{saka+08oe,xia+08oe,yang+23amr,zaie+23jpca}, providing a versatile platform for systematically tuning both conjugation length and geometry~\cite {mish+09cr}.
This structural and electronic flexibility is intimately connected with the characteristic symmetries exhibited by thiophene oligomers in their idealized, flat conformation. The single thiophene ring (hereafter denoted as 1T) is a non-centrosymmetric molecule belonging to the $C_{2v}$ group~\cite{cumm+24cej} and representative of odd-membered thiophene oligomers. On the other hand, quaterthiophene (4T) with its even number of rings, is a centrosymmetric member of the $C_{2h}$ point group~\cite{belj+93jcp,rubi+05chpch}.

Exciting thiophene oligomers with strong electric fields triggers several intriguing phenomena~\cite{greb+95jpc,lanz+01jppa,vare+12prl}. In recent work~\cite{drio+25jpcl}, we demonstrated that oligothiophenes can host optical limiting, a third-order nonlinearity consisting of the frequency-selective attenuation of transmission upon intense irradiation~\cite{tutt-bogg93pqe}. By complementing real-time time-dependent density-functional theory (RT-TDDFT) simulations within the Yabana-Bertsch scheme~\cite{yaba-bert96prb,cocc+14prl,guan+21pccp} with excitations driven by resonant pulses~\cite{krum+20jcp}, we could attribute this behavior primarily to excited-state absorption (ESA)~\cite{drio+25jpcl}. By tracking single-particle excited populations under resonant driving, we provided direct evidence of state-to-state transitions from the pumped bright state ($S_{\mathrm B}$) into higher-lying excited manifolds ($S_n$) under intense field conditions~\cite{drio+25jpcl}.

The non-perturbative nature of RT-TDDFT offers a powerful, cost-effective \textit{ab initio} framework to model strong-field dynamics in large conjugated molecules without truncating perturbative expansions~\cite{degi+13chpch,goin-lest18wires,ghos-roy22cpl,vang-snij98jcp,park+18jctc}. However, standard orbital projections used to interpret these numerical results often lack the resolution that is needed to untangle complex excitation pathways. For instance, high-harmonic generation (HHG) serves as a sensitive probe for ultrafast electronic dynamics~\cite{marc+25pccp}, where point-group symmetry dictates selection rules for even- and odd-order harmonics. However, isolating the correlation between transient population transfers and nonlinear spectral signatures remains challenging. In particular, identifying robust analytical bridges to decipher the multi-configurational electronic states encoded within real-time density matrices remains a crucial theoretical task~\cite{guan+21pccp}.

In this work, we investigate 1T and 4T, taken as representatives of odd- and even-membered thiophene oligomers, subject to resonant pulses of increasing intensity. By combining RT-TDDFT with a many-body non-perturbative population framework, which coincides with conventional estimates in the weak-field limit, we disentangle the coupled roles of $\pi$-conjugation length and point-group symmetry in shaping the nonlinear optical response across weak, intermediate, and strong-field regimes. We demonstrate that extending the backbone length in 4T reduces the intensity threshold for nonlinear absorption by orders of magnitude compared to 1T, driving rapid ground-state depletion. State-resolved dynamic projections provide direct evidence of transient channel switching into higher-lying excited states ($S_{\mathrm B}\to S_n$), establishing the microscopic onset of field-induced ESA. Finally, we show how molecular symmetry governs HHG for a field polarized along the molecular axis: inversion symmetry in 4T completely quenches even-order harmonics, while strong spatial anisotropy along the conjugated backbone enhances the signal and lowers the threshold for high-order emission.

\section{Methodology}
\label{sec:methods}

\subsection{Ab initio Computational Framework}

RT-TDDFT provides a first-principles, non-perturbative propagation scheme for the electron density subject to external time-dependent fields~\cite{rung-gros84prl,cast+04jcp}. The time evolution of the Kohn-Sham (KS) states, $\phi_i(r,t)$, is governed by the time-dependent KS (TDKS) equations:
\begin{equation}
i\hbar \frac{\partial}{\partial t} \phi_i(r,t) =
\left[
-\frac{\hbar^2}{2m}\nabla^2 + V_{\mathrm{eff}}\left[ \rho \right](r,t) - e\,r\mathcal{E}(t)
\right]\phi_i(r,t),
\label{eq:TDKS}
\end{equation}
where 
\begin{equation}
\rho(r,t)=\sum_{i=1}^{N_{el}} f_i|\phi_i(r,t)|^2
\end{equation}
is the time-dependent electron density,  $N_{el}$ is the total number of (valence) electrons, and $0<f_i<1$ is the spin-degenerate initial occupation, such that $\int \rho(r,t)\, d\mathbf{r} = \sum_i f_i = N_{el}$. The effective potential $V_{\mathrm{eff}}[\rho](\mathbf{r},t) = V_{\mathrm{ext}}(\mathbf{r}) + V_{\mathrm{H}}[\rho](\mathbf{r},t) + V_{\mathrm{xc}}[\rho](\mathbf{r},t)$ contains the external ionic, Hartree, and exchange-correlation contributions, respectively. Light-matter interaction is evaluated in the length gauge within the electric dipole approximation, where $\mathcal{E}(t)=\mathcal{E}_0 f(t)
\cos[\omega(t-t_0)+\varphi]\hat{\mathbf{e}}$ represents the linearly polarized driving pulse with amplitude $\mathcal{E}_0$, envelope $f(t)$, carrier frequency $\omega$, phase $\varphi$, and polarization direction $\hat{\mathbf{e}}$.

The dynamic electronic response is monitored via the time-dependent dipole moment along the direction $r_i \in \{ x,y,z \}$:
\begin{equation}
\mu_i(t) = -e \int r_i\, \rho(r,t)\, d\mathbf{r},
\end{equation}
from which HHG spectra are calculated via the Fourier transform of the dipole acceleration autocorrelation function~\cite{yaba-bert96prb,krum+20jcp}:
\begin{equation}
S(\omega) \propto \sum_i \left| \int \ddot{\mu}_i(t)\, e^{i\omega t}\, dt \right|^2.
\end{equation}
We adopt the fixed-nuclei approximation, as we focus on ultrafast electronic dynamics below 30-fs timescales in which the ionic motion does not evolve significantly. 

\subsection{Nonlinear Response and Population Analysis}

To characterize the efficiency of nonlinear absorption, we monitor the total energy uptake from the initial time $t_i$ (pre-pulse) and the final time $t_f$ (post-pulse):
\begin{equation}
\Delta E(t_f)=E_{\mathrm{tot}}(t_f)-E_{\mathrm{tot}}(t_i) = \int_{t_i}^{t_f} \dot d_\parallel(t)\, \mathcal{E}_\parallel(t)\, dt,
\label{eq:energy_uptake}
\end{equation}
where $d_\parallel(t)$ and $\mathcal{E}_\parallel(t)$ are the dipole and electric-field components projected along the laser-polarization direction. The dipole work integral in Eq.~\eqref{eq:energy_uptake} serves as an internal consistency check on $\Delta E$. The effective number of absorbed photons is $N_{\mathrm{ph}} = \Delta E / (\hbar\omega)$, where $\hbar\omega$ is the carrier photon energy, whereas the total count of promoted electrons, $N_{\mathrm{ex}}(t)$, is evaluated from the population loss in the initially occupied reference KS manifold:
\begin{equation}
N_{\mathrm{ex}}(t) = N_{\mathrm{el}} - \sum_{n\in\mathrm{occ}} p_n(t). 
\label{eq:Nex}
\end{equation}
Here, $p_n(t) = \sum_{i\in\mathrm{occ}} f_i^0 |\langle\phi_n^0|\phi_i(t)\rangle|^2$ is the electron-weighted occupation of reference level $n$, and $f_i^0$ is the initial ground-state occupation.
To resolve where the electronic population is distributed in energy, we construct the time-dependent occupation density $N(\varepsilon,t)$ by broadening $p_n(t)$ over the field-free orbital energies $\varepsilon_n^0$ using normalized Gaussians $G_n(\varepsilon)$:
\begin{equation}
N(\varepsilon,t) = \sum_n p_n(t)\, G_n(\varepsilon).
\label{eq:occupation_density}
\end{equation}
The reference density of states (DOS) is constructed analogously as $D(\varepsilon) = \sum_n w_n^0 G_n(\varepsilon)$, with weights $w_n^0 = f_n^0$ if the state $n$ is occupied, and $w_n^0 = 1$ if $n$ is a virtual orbital.

While $N_{\mathrm{ex}}(t)$ and $N(\varepsilon,t)$ quantify charge promoted out of the occupied manifold, single-particle diagnostics cannot assign that charge to specific many-electron excited configurations. To obtain bounded many-body transition probabilities, we evaluate the time-dependent Slater determinant $|\Phi(t)\rangle$ constructed from the occupied TDKS orbitals. We partition the reference-to-propagated overlap matrix $C(t)=\langle\phi_p^0|\phi_n(t)\rangle$ into occupied and virtual blocks:
\begin{align}
& C(t)=
\begin{pmatrix}
A(t)\\
B(t)
\end{pmatrix} \\
& A_{in}(t)=\langle\phi_i^0|\phi_n(t)\rangle, \\
& B_{an}(t)=\langle\phi_a^0|\phi_n(t)\rangle .
\end{align}
The survival probability of the ground-state determinant is $P_0(t) = |\langle\Phi_0|\Phi(t)\rangle|^2 = |\det A(t)|^2$. Through the exact Thouless representation~\cite{thou60np,ring-schu80}, the propagated determinant is expressed as
\begin{equation}
|\Phi(t)\rangle = \det A(t) \exp\!\left[\sum_{ia} Z_{ai}(t) c_a^\dagger c_i\right] |\Phi_0\rangle, 
\label{eq:thouless_main}
\end{equation}
where $Z(t)= B(t)A^{-1}(t)$ encodes occupied-to-virtual orbital rotations.
Projecting single excitations onto Casida's linear-response vectors~\cite{casi95}, $|\Phi_I\rangle=\sum_{ia}X_{ia}^{(I)}c_a^\dagger c_i|\Phi_0\rangle$, yields state-resolved populations:
\begin{equation}
P_I(t)=|\langle\Phi_I|\Phi(t)\rangle|^2=\left|\sum_{ia}X_{ia}^{(I)*}\det\!\bigl(A(t)\big|_{i\to a}\bigr)\right|^2.
\label{eq:PI_manybody}
\end{equation}
Using Cramer's rule ($\det(A|_{i\to a}) = \det A \cdot Z_{ai}$), this reduces to:
\begin{equation}
P_I(t)=P_0(t)\,|\mathcal{Z}_I(t)|^2,
\qquad
\mathcal{Z}_I(t)=\sum_{ia}X_{ia}^{(I)*}Z_{ai}(t).
\label{eq:PThou_main}
\end{equation}
This form requires only a single $\mathcal{O}(N_{\mathrm{occ}}^3)$ matrix inversion per time step. By contrast, conventional density-matrix estimators $P_I^{\mathrm{CIS}}(t) = \sum X_{ia}^{(I)*} X_{jb}^{(I)} K_{ia,jb}(t)$ omit the ground-state depletion renormalization $A^{-1}(t)$, coinciding with Eq.~\eqref{eq:PThou_main} only in the weak-field limit (see Supplementary Material, Sec.~S1). We reserve uppercase $P_s$ for determinant-based probabilities and lowercase $p_s$ for single-particle occupations.

\subsection{Coupling-Strength Parameter and Excitation Regimes}

To compare oligomers with different transition dipole moments and excitation energies on an equal footing, we introduce the dimensionless dipolar coupling parameter
\begin{equation}
\eta=
\frac{|\mathcal{E}_0\mu_\parallel|}{\hbar\omega}
\label{eq:eta}
\end{equation}
to classify the excitation strength. In Eq.~\eqref{eq:eta},  $\mu_\parallel$ is the transition dipole moment of the pumped bright state $S_B$, oriented parallel to the pulse polarization. 

We define two threshold criteria to partition the light--matter interaction into operational regimes (Table~\ref{tab:eta_regimes}). The nonlinear threshold ($\eta_{\mathrm{nl}}$) marks the onset of nonlinear but perturbative response, defined when the peak population $P_{S_{\mathrm B}}^{\max}(\eta) = \max_t P_{S_{\mathrm B}}(t)$ of the pumped bright state $S_{\mathrm B}$ deviates by more than 10\% from the linear-response (single-photon) scaling ($P_{S_{\mathrm B}}^{\max}(\eta) \propto \eta^2 \propto I$):
    \begin{equation}
    \eta_{\mathrm{nl}} = \min \left\{ \eta : \left| \frac{P_{S_{\mathrm B}}^{\max}(\eta)}{a_{\mathrm B}\eta^2} - 1 \right| > 0.1 \right\} \sim 10^{-2},
    \label{eq:eta_nl}
    \end{equation}
    where $a_{\mathrm{B}}$ is fitted in the low-field linear regime.
At higher field strengths, the non-perturbative threshold ($\eta_{\mathrm{np}}$) signals the emergence of excitation-order mixing beyond the single-excitation manifold. From the excitation-rank decomposition of the propagated determinant $|\Phi(t)\rangle$, the total single-excitation population is $P_1^{\mathrm{tot}}(t) = P_0(t)\sum_m \lambda_m(t)$, where $\{\lambda_m\}$ are eigenvalues of $Z(t)Z^\dagger(t)$. The residual population leaking into double- and higher-excitation ranks is given by
    \begin{equation}
    P_{\geq2}(t) = 1 - P_0(t) - P_1^{\mathrm{tot}}(t).
    \label{eq:Pge2}
    \end{equation}
    We define $\eta_{\mathrm{np}}$ as the point where the peak higher-rank population $P_{\geq2}^{\max} = \max_t P_{\geq2}(t)$ reaches 0.1:
    \begin{equation}
    \eta_{\mathrm{np}} = \min \left\{ \eta : P_{\geq2}^{\max}(\eta) \ge 0.1 \right\} \sim 10^{-1}.
    \label{eq:eta_np}
    \end{equation}
    A value of $P_{\geq2}^{\max} \ge 0.1$ indicates that over $10\%$ of the propagated state norm lies outside the ground-plus-singles manifold, rendering singles-only population estimators quantitatively unreliable.

To further monitor intensity scaling trends, local power-law exponents $\alpha^{(Q)} = d\log Q / d\log I$ for post-pulse energy uptake ($\Delta E$) and excited electron count ($N_{\mathrm{ex}}$) are extracted along intensity scans as visual diagnostics of population saturation ($\alpha < 1$) or multi-photon absorption ($\alpha > 1$), see section S1.4 and Table S2 in the Supplementary Material.

\begin{table*}[t!]
\caption{Operational excitation regimes. The boundaries are approximate
order-of-magnitude scales: $\eta_{\mathrm{nl}}\sim10^{-2}$ and
$\eta_{\mathrm{np}}\sim10^{-1}$.}
\label{tab:eta_regimes}
\begin{ruledtabular}
\renewcommand{\arraystretch}{1.35}
\begin{tabular}{lll}
Regime & Coupling range & Physical meaning \\
\hline
weak / linear
&
$\eta\lesssim\eta_{\mathrm{nl}}$
&
\parbox[t]{0.42\linewidth}{linear response; CIS/Casida singles picture valid}
\\
intermediate / nonlinear
&
$\eta_{\mathrm{nl}}\lesssim\eta\lesssim\eta_{\mathrm{np}}$
&
\parbox[t]{0.42\linewidth}{saturation/ESA; ground state and single excitations dominate}
\\
strong / non-perturbative
&
$\eta\gtrsim\eta_{\mathrm{np}}$
&
\parbox[t]{0.42\linewidth}{excitation-order mixing; leakage beyond singles is significant}
\end{tabular}
\end{ruledtabular}
\end{table*}

\subsection{Computational Details}

All RT-TDDFT calculations performed in this work were carried out using the code \textsc{Octopus}~\cite{tanc+20jcp}, implementing real-space grids. Ground-state electronic structures were evaluated within the local-density approximation using the Perdew--Zunger functional~\cite{perd-zung81pr} and norm-conserving Hartwigsen--Goedecker--Hutter pseudopotentials~\cite{hart+98prb}. Real-space integrations were discretized on a uniform grid with a spacing of $0.18$~\AA{} enclosed within a spherical simulation box of radius $6.0$~\AA{}. Molecular geometries were optimized using the FIRE algorithm~\cite{bitz+06prl} to a maximum force threshold of $10^{-4}$~eV/\AA{}.

The TDKS equations were propagated using the enforced time-reversal symmetry algorithm~\cite{cast+04jcp} with a time step of 3~as. Resonant optical driving was modeled using linearly polarized $\cos^2$-enveloped pulses aligned with the transition dipole of the bright $\pi\to\pi^*$ state ($S_B$):
\begin{align}
\mathcal{E}(t)&=\mathcal{E}_0 f(t)
\cos[\omega(t-t_0)+\varphi],
\label{eq:pulse_field}\\
f(t)&=
\begin{cases}
\cos^2\!\left[\dfrac{\pi(t-t_0)}{2T_p}\right], & |t-t_0|\le T_p,\\
0, & |t-t_0|>T_p,
\end{cases}
\label{eq:pulse_envelope}
\end{align}
where $T_p$ is the half-duration of the envelope. Carrier frequencies were set to linear absorption maxima: $\hbar\omega = 5.5~\mathrm{eV}$ for 1T ($2T_p = 14.2~\mathrm{fs}$) and $2.5~\mathrm{eV}$ for 4T ($2T_p = 28.4~\mathrm{fs}$). Transition dipoles are $\mu_\parallel = 0.384~\text{\AA}$ (1T) and $2.135~\text{\AA}$ (4T).

\section{Results and Discussion}
\label{sec:results}

\subsection{Energy Uptake and Intensity-Dependent Excitation Dynamics}
\label{subsec:energy_uptake}

\begin{figure*}
    \centering
    \includegraphics[width=1\linewidth]{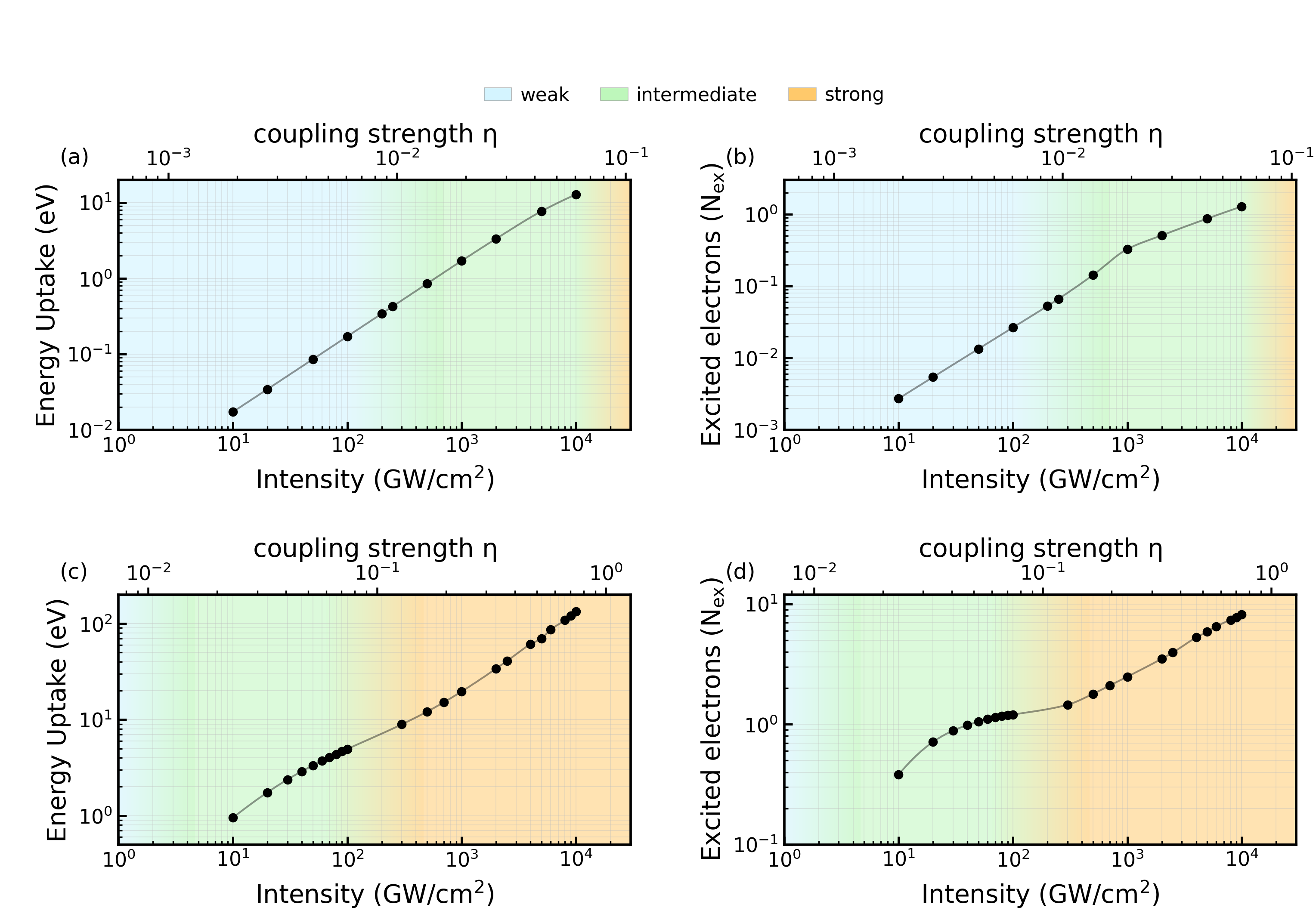}
    \caption{Post-pulse energy uptake and number of excited electrons as functions of peak laser intensity (lower axis) and the dimensionless coupling strength $\eta$ (upper axis) for (a)-(c) 1T (b)-(d) 4T, respectively. Shaded backgrounds identify the weak/linear (light blue), intermediate/nonlinear (light green), and strong/non-perturbative (orange) excitation regimes, with blended transitions around $\eta_{\mathrm{nl}}\sim10^{-2}$ and $\eta_{\mathrm{np}}\sim10^{-1}$. }
    \label{fig:Energy_Uptake}
\end{figure*}

We start our analysis of the intensity-dependent electronic response of the thiophene oligomers by inspecting the total post-pulse energy uptake and number of excited electrons at increasing field strengths (Fig.~\ref{fig:Energy_Uptake}).  At low coupling ($\eta \lesssim 10^{-2}$, blue shaded region), 1T exhibits linear scaling ($\alpha \approx 1.0$) for both $\Delta E$ and $N_{\mathrm{ex}}$ [Figs.~\ref{fig:Energy_Uptake}(a,b)], characteristic of perturbative single-photon absorption. For 1T, energy uptake remains nearly linear up to multi-$\mathrm{TW\,cm^{-2}}$ intensities ($\eta \simeq 4.3 \times 10^{-2}$ at $5~\mathrm{TW\,cm^{-2}}$), before moderately bending sublinearly due to ground-state depletion and absorption saturation~\cite{drio+25jpcl}. In contrast, $N_{\mathrm{ex}}$ is far more sensitive to field strength: it begins to flatten around $500~\mathrm{GW\,cm^{-2}}$ ($\eta \simeq 1.4 \times 10^{-2}$), marking the onset of the intermediate nonlinear regime where additional field energy no longer produces excited electrons at a constant rate.

Extending the $\pi$-conjugated backbone in 4T shifts these excitation regimes to substantially lower intensities due to its larger transition dipole moment. Even at the lowest sampled intensity of $10~\mathrm{GW\,cm^{-2}}$, 4T already sits at $\eta \simeq 2.3 \times 10^{-2}$, entering the intermediate regime [Figs.~\ref{fig:Energy_Uptake}(c,d)]. The energy uptake $\Delta E$ for 4T increases smoothly across the intensity range, exceeding $100~\mathrm{eV}$ above $1~\mathrm{TW\,cm^{-2}}$ ($\eta > 2.3 \times 10^{-1}$). 
However, $N_{\mathrm{ex}}$ in 4T displays a distinct two-step behavior [Fig.~\ref{fig:Energy_Uptake}(d)]: $N_{\mathrm{ex}}$ initially flattens into a pronounced plateau near $N_{\mathrm{ex}} \approx 1.0$ around $50\text{--}100~\mathrm{GW\,cm^{-2}}$ ($\eta \approx 5\text{--}7 \times 10^{-2}$), signaling complete single-electron population saturation of the resonant bright state. Upon crossing into the strong/non-perturbative regime ($\eta \gtrsim \eta_{\mathrm{np}} \sim 10^{-1}$, $I \gtrsim 180~\mathrm{GW\,cm^{-2}}$), $N_{\mathrm{ex}}$ undergoes a secondary superlinear growth, exceeding $N_{\mathrm{ex}} \approx 8$ at $10~\mathrm{TW\,cm^{-2}}$. 

This decoupling between the slopes of $\Delta E$ and $N_{\mathrm{ex}}$ reveals that under intense driving, incoming field energy is no longer consumed by generating additional single-particle excitations from the ground state. Instead, energy is absorbed through secondary transitions and population redistribution within the already populated excited-state manifold. This mechanism, which is more pronounced in 4T due to its extended $\pi$-system, higher excited-state density, and stronger transition dipole couplings between excited states, motivates the state-resolved population dynamics and energy-resolved occupation density analyses detailed in the following sections.

\subsection{High-Harmonic Generation and Anisotropic Nonlinear Response}
\label{subsec:hhg}

\begin{figure*}[t]
    \centering
    \includegraphics[width=\linewidth]{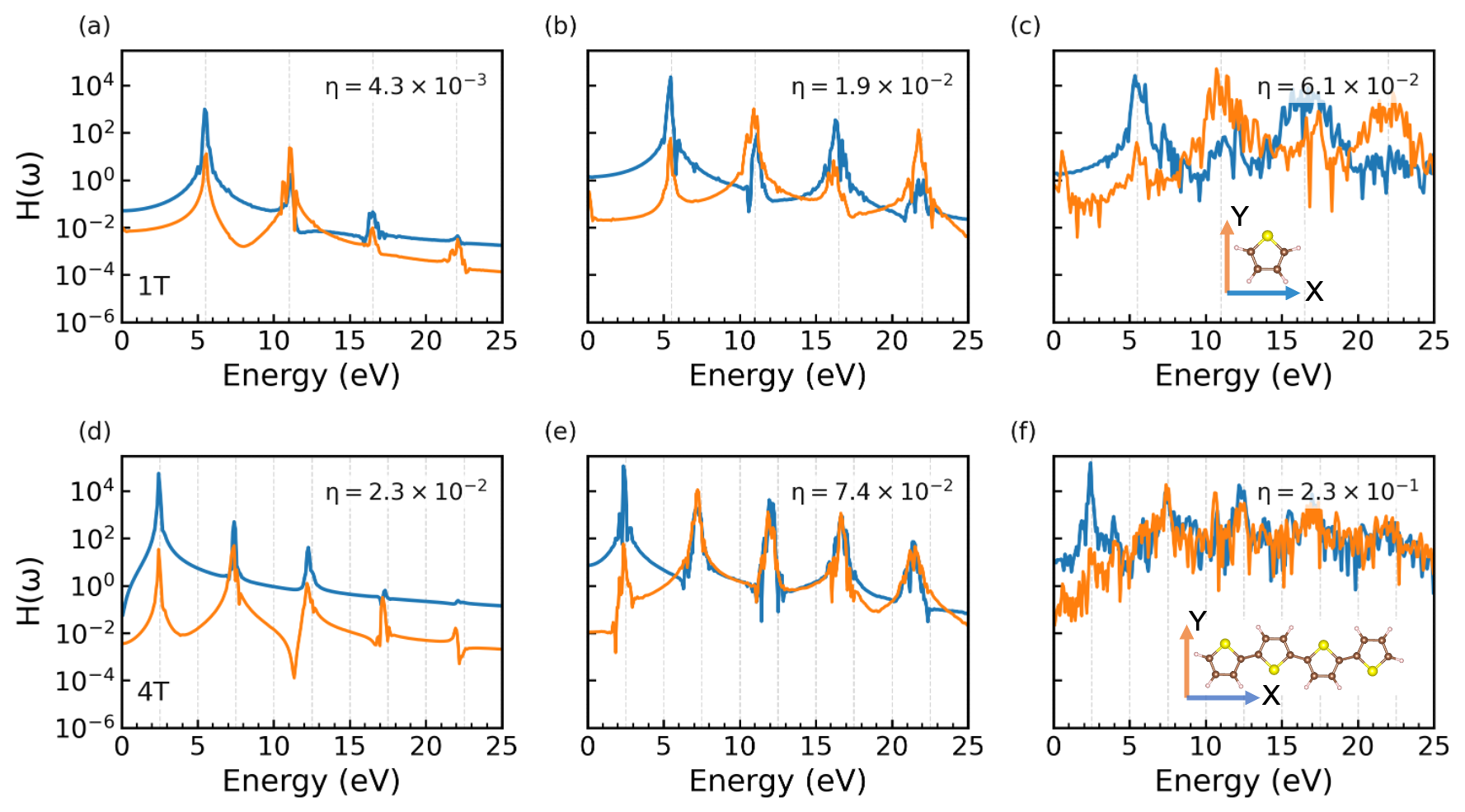}
    \caption{HHG spectra of (a)-(c) 1T and (d)-(f) 4T across increasing coupling strengths $\eta$ under $x$-polarized pump. Blue and orange curves represent harmonic emission along the $x$- and $y$-axes, see insets in panels (c) and (f). Vertical dotted lines indicate integer multiples of the fundamental driving frequency $\omega$.}
    \label{fig:HHG}
\end{figure*}

To probe the dynamic hyperpolarizability and symmetry constraints of oligothiophenes, we examine the HHG spectra of 1T and 4T under an $x$-polarized pulse of increasing intensity, considering emission parallel to both molecular axes (Fig.~\ref{fig:HHG}). 
For the single thiophene ring, which belongs to the non-centrosymmetric $C_{2v}$ point group, the lack of spatial inversion symmetry relaxes electric-dipole selection rules. Consequently, both odd and even harmonics are dipole-allowed [Fig.~\ref{fig:HHG}(a)--(c)]. Notably, while odd harmonics are emitted predominantly along the driving field axis $x$ (blue curves), even-harmonic peaks, particularly  $2\omega$ ($11~\mathrm{eV}$) and $4\omega$ ($22~\mathrm{eV}$), appear with pronounced intensity along the transverse direction ($y$, orange curves). This transverse even-harmonic generation stems from the permanent ground-state dipole moment of thiophene, which is directed along its $C_2$ symmetry axis ($y$), creating a cross-polarized nonlinear response.
 In contrast, planar 4T ($C_{2h}$ point-group symmetry) possesses an inversion center that enforces parity selection rules. Even-harmonic pathways are therefore completely extinguished and high-frequency emission is restricted exclusively to odd multiples of the driving frequency along both the longitudinal (blue) and transverse (orange) emission directions [Fig.~\ref{fig:HHG}(d)--(f)]. 

Beyond selection rules, the HHG spectra illustrate a strong spatial anisotropy in 4T. At low-to-intermediate coupling ($\eta = 2.3 \times 10^{-2}$), harmonic yield along the long axis ($x$, blue curves) is orders of magnitude stronger at lower harmonic orders than in the transverse direction ($y$, orange), confirming that the primary hyperpolarizability is driven by electron delocalization along the thiophene backbone. Comparing the two oligomers, 4T enters the nonlinear harmonic regime at field strengths nearly two orders of magnitude lower than 1T, mirroring the behavior in $N_{\mathrm{ex}}$ discussed in Sec.~\ref{subsec:energy_uptake}, and confirming that $\pi$-conjugation extension amplifies both the instantaneous optical response and the field-induced population transfer. 

At the highest driving fields [$\eta \sim 10^{-1}$, Figs.~\ref{fig:HHG}(c,f)], the discrete harmonic peaks broaden into a multi-peak continuum background. This spectral distortion originates from rapid, non-adiabatic phase modulation driven by massive transient population shifts within the excited-state manifold, providing a natural bridge to analyze the state-resolved populations in the following section.

\subsection{Population Dynamics and Non-Perturbativity Analysis}
\label{subsec:pop_dyn}

\begin{figure*}
    \centering
    \includegraphics[width=\linewidth]{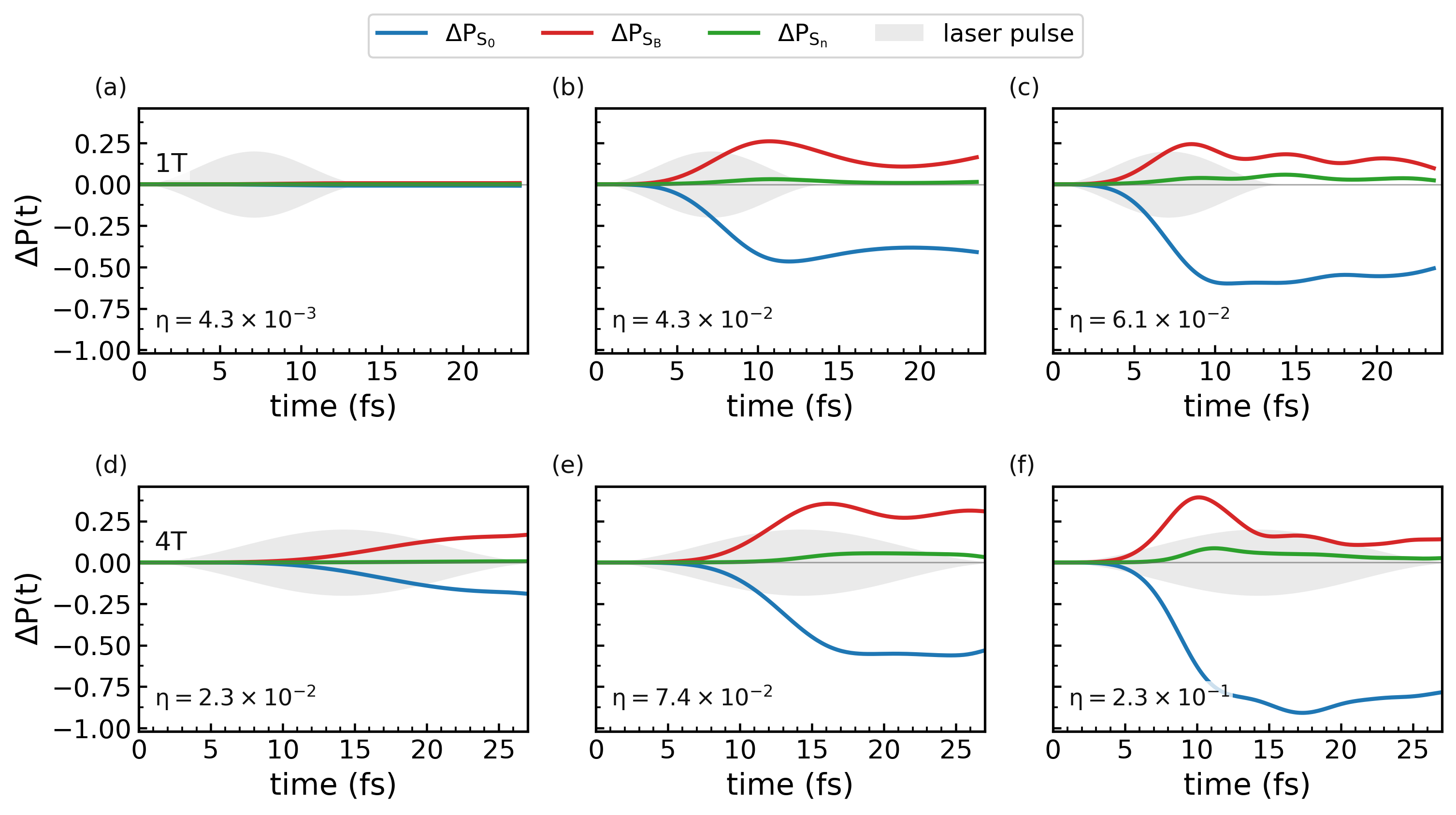}
    \caption{State-resolved transient population dynamical variations of the ground state ($\Delta P_{\mathrm{GS}}(t)$, blue), the pumped bright-state band ($\Delta P_{S_{\mathrm B}}(t)$, red), and the higher excited-state absorption band ($\Delta P_{S_n}(t)$, green) in (a)-(c) 1T and (d)-(f) 4T across increasing coupling strengths $\eta$. The $S_B$ band contains the resonant bright peak at 5.7~eV for 1T and at 2.5~eV for 4T. The $S_n$ band collects higher-lying Casida states in the 7.02-9.06~eV window for 1T and 3.58-4.09~eV for 4T. The gray shaded area denotes the laser-pump envelope. }
    \label{fig:Population}
\end{figure*}

To unravel the microscopic electronic mechanisms governing the nonlinear absorption thresholds, we track the instantaneous many-body state populations. By projecting the propagated TDKS determinant onto Casida-resolved excited-state manifolds, we unveil the temporal competition between ground-state depletion, the pumped bright state, and higher-lying absorption channels under resonant laser driving (Figure~\ref{fig:Population}).

Across both oligomers, increasing field strength transforms the population dynamics through the different excitation regimes (Table~\ref{tab:eta_regimes}). At low coupling [Figure~\ref{fig:Population}(a),(d)], light-matter interaction operates in the weak, linear perturbative limit: ground-state depletion is minimal ($\Delta P_{\mathrm{GS}} > -0.01$), driving a weak and smooth population transfer into the pumped bright state ($\Delta P_{S_{\mathrm B}}$), while higher-lying states ($\Delta P_{S_n}$) remain essentially unpopulated. At intermediate coupling [Figure~\ref{fig:Population}(b),(e)], ground-state depletion becomes pronounced and $\Delta P_{S_B}$ approaches saturation. Near the pulse maximum, $\Delta P_{S_B}(t)$ exhibits coherent Rabi-like oscillations before stabilizing at a residual population plateau after the pulse elapses. Under strong driving [Figure~\ref{fig:Population}(c),(f)], ground-state depletion is substantial, reaching $\Delta P_{\mathrm{GS}}=-0.6$ in 1T and $\Delta P_{\mathrm{GS}}=1.0$ in 4T, while $\Delta P_{S_B}(t)$ approaches a transient maximum near the pulse peak before decaying. 

This non-monotonic behavior of $S_B$, together with increasing $\Delta P_{S_n}$, provides clear evidence of ESA, where population initially transferred to $S_B$ is subsequently promoted into higher-lying manifolds. Furthermore, the population mismatch between ground-state depletion ($-\Delta P_{\mathrm{GS}}$) and the sum of tracked singles ($\Delta P_{S_{\mathrm B}} + \Delta P_{S_n}$) indicates that population weight is leaving the single-excitation manifold.
Extending the $\pi$-conjugation length in 4T lowers the field threshold for activating the $S_{\mathrm B}\to S_n$ channel by orders of magnitude compared to 1T while preserving these qualitative features. This macroscopic shift in channel activation underpins both the lower saturation threshold in energy uptake and the spectral broadening in HHG.

The high-coupling population mismatch marks the departure from a single-excitation-dominated picture. To quantify this crossover without relying solely on visual state analysis, we evaluate the time-dependent non-perturbativity descriptor $P_{\geq2}(t)$ [Eq.~\eqref{eq:Pge2}], which assesses the total determinant weight lying outside the ground-state plus complete single electron--hole sector. In our TDKS framework, this weight arises from the antisymmetrized product structure of the occupied orbital rotations, rather than explicit multi-reference correlation.

\begin{figure*}
    \centering
    \includegraphics[width=\linewidth]{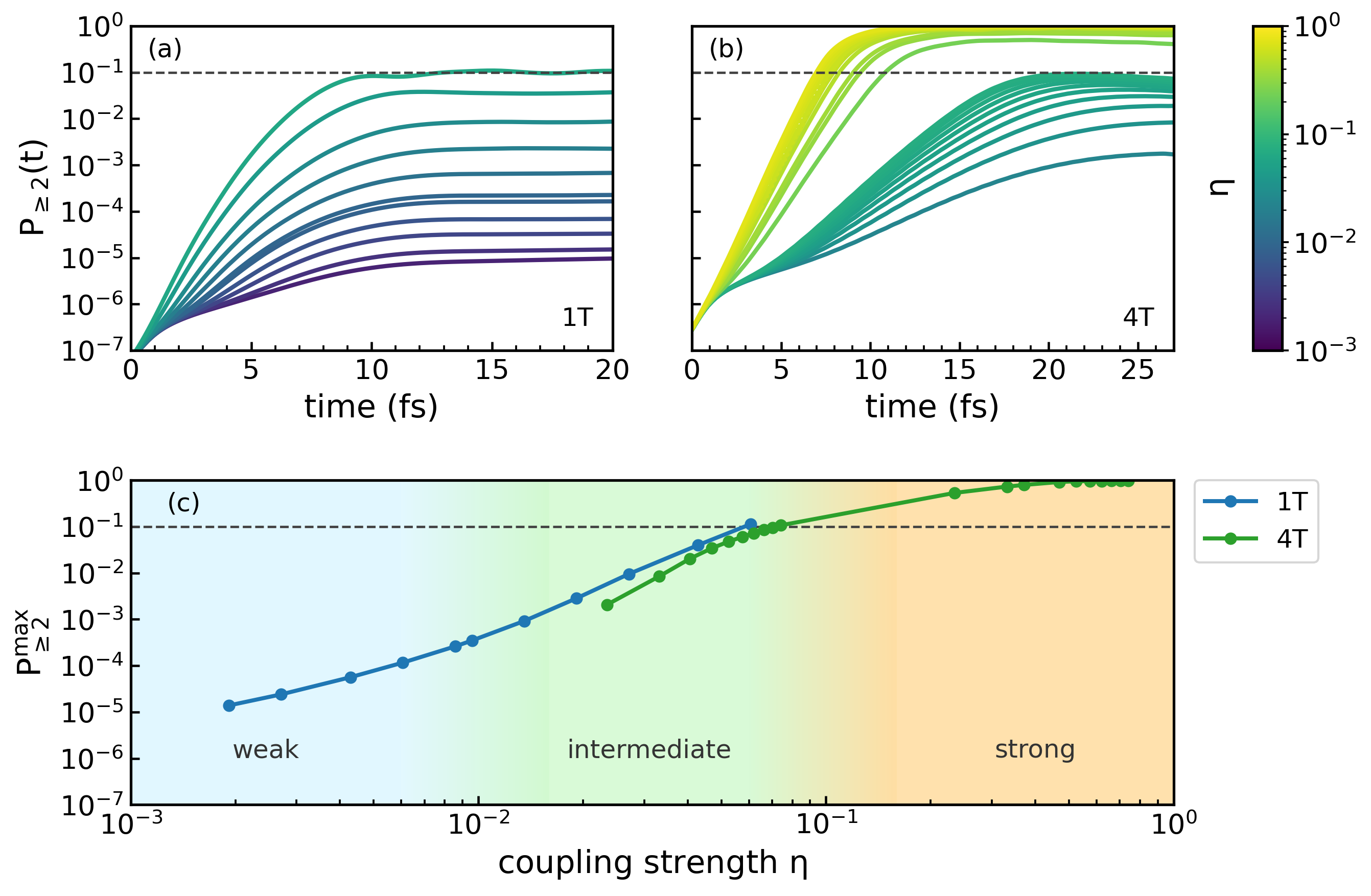}
    \caption{Non-perturbativity descriptor $P_{\geq2}(t)$ for (a) 1T and (b) 4T across the explored coupling-strength range, with trajectories colored according to the value of $\eta$. (c) Maximum value $P_{\geq2}^{\max}=\max_tP_{\geq2}(t)$ as a function of $\eta$ for 1T (blue) and 4T (green), with the horizontal dashed line marking the conservative crossover criterion $P_{\geq2}=0.1$.}
    \label{fig:Pge2_time}
\end{figure*}

As shown in Fig.~\ref{fig:Pge2_time}(a), $P_{\geq2}(t)$ in 1T remains negligible throughout the weak and intermediate regimes, reaching $P_{\geq2}^{\max}=4.00\times10^{-2}$ at $5~\mathrm{TW\,cm^{-2}}$ ($\eta \simeq 4.3 \times 10^{-2}$). Only at the highest intensity ($I=10~\mathrm{TW\,cm^{-2}}$, $\eta \simeq 6.1 \times 10^{-2}$), 1T crosses the $10\%$ threshold, with $P_{\geq2}^{\max}=0.113$ at $t\simeq13.1~\mathrm{fs}$. Thus, in 1T, substantial ground-state depletion and saturation occur prior to significant higher-rank mixing.
The situation is markedly different for 4T [Fig.~\ref{fig:Pge2_time}(b)], where the threshold is overcome already at $I= 90~\mathrm{GW\,cm^{-2}}$ ($P_{\geq2}^{\max} = 0.096$) and crosses it at $I=100~\mathrm{GW\,cm^{-2}}$ ($\eta \simeq 7.4 \times 10^{-2}$, $P_{\geq2}^{\max} = 0.107$). At $1~\mathrm{TW\,cm^{-2}}$ ($\eta \simeq 2.3 \times 10^{-1}$), $P_{\geq2}^{\max}$ reaches $0.534$, crossing the threshold early in the pulse ($t \simeq 10.9~\mathrm{fs}$), and approaches unity at higher fields.

The peak-value summary in Fig.~\ref{fig:Pge2_time}(c) confirms that $\eta$ serves as a unified control parameter across different oligomer lengths. In both 1T and 4T, $P_{\geq2}^{\max}$ crosses $0.10$ near $\eta \sim 10^{-1}$ ($\eta \approx 0.07$ for 4T and $\eta \approx 0.06$ for 1T). This alignment demonstrates that the lower nonlinear threshold in 4T is not merely an increase in  $N_{\mathrm{ex}}$, but represents the onset of higher-order mixing driven by enhanced light-matter coupling.

\subsection{Energy-Resolved Occupation Density and Orbital Redistribution}
\label{subsec:occ_den}

\begin{figure*}
    \centering
    \includegraphics[width=\linewidth]{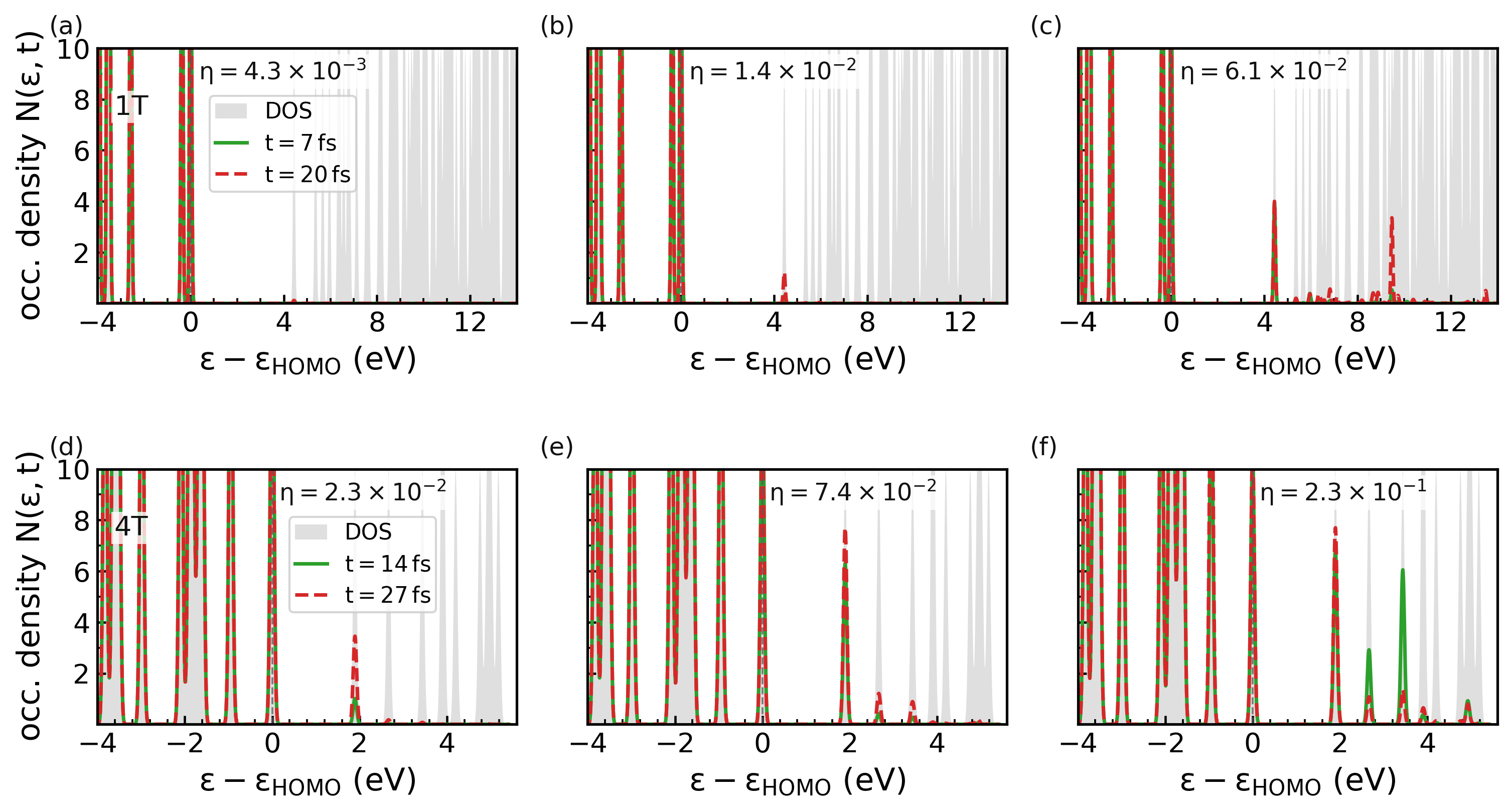}
\caption{Time-dependent occupation density $N(\varepsilon,t)$ for (a)-(c) 1T and (d)-(f) 4T at weak, intermediate, and strong coupling strengths. Solid green and dashed red curves denote the projected electronic distribution at the pulse-envelope maximum ($t=7\,\text{fs}=t_0$) and after pulse completion ($t=20\,\text{fs}=t_f$), respectively. The gray shaded area is the field-free KS density of states (DOS), broadened with normalized Gaussian functions of width $\sigma=0.04$~eV. Energies are shifted with respect to the highest-occupied molecular orbital (HOMO) set at 0~eV.}
    \label{fig:occ_density}
\end{figure*}

To gain deeper microscopic insight into the redistribution of the absorbed energy within the molecular orbital manifold, we analyze the energy-resolved occupation density $N(\varepsilon,t)$ [Eq.~\eqref{eq:occupation_density}] at the pulse peak ($t=t_0$, green curves) and the residual distribution after pulse completion ($t=t_f$, dashed red curves) with the ground-state density of states (DOS, grey areas) across weak-, intermediate-, and strong-field regimes )Figure~\ref{fig:occ_density}). This analysis follows from the observation above that $N_{\mathrm{ex}}$ saturates while $\Delta E$ continues to increase, suggesting that further energy uptake is associated with redistribution of the excited population among higher-energy KS levels rather than only with the creation of additional excited electrons.

In 1T, weak coupling [$\eta = 4.3 \times 10^{-3}$, Fig.~\ref{fig:occ_density}(a)] induces negligible population transfer, with $N(\varepsilon,t)$ matching the unperturbed ground-state DOS. Increasing the field to intermediate coupling [$\eta = 1.4 \times 10^{-2}$, Fig.~\ref{fig:occ_density}(b)] drives permanent population transfer into the lowest unoccupied level near 4.5~eV, accompanied by slight depletion of valence states.
Under strong-field conditions [$\eta = 6.1 \times 10^{-2}$, Fig.~\ref{fig:occ_density}(c)], deep valence states down to -4.0~eV suffer substantial depletion. Simultaneously, electrons are promoted beyond the LUMO into higher virtual levels near $+9.5~\mathrm{eV}$. The close overlap between transient peak occupation (green) and residual post-pulse population (red) confirms that this high-energy redistribution represents permanent excitation.

In 4T, the denser DOS facilitates step-wise excitation pathways across the $\pi^*$-conduction manifold at substantially lower field strengths. At low coupling [$\eta = 2.3 \times 10^{-2}$, Fig.~\ref{fig:occ_density}(d)], excitation predominantly promotes electrons from the highest-occupied to the lowest-unoccupied orbital. At intermediate coupling [$\eta = 7.4 \times 10^{-2}$, Fig.~\ref{fig:occ_density}(e)], population spreads across multiple  higher-lying virtual levels at $+2.8~\mathrm{eV}$ and $+3.5~\mathrm{eV}$, alongside valence-band depletion. At strong coupling [$\eta = 2.3 \times 10^{-1}$, Fig.~\ref{fig:occ_density}(f)], $N(\varepsilon,t)$ shows deep valence evacuation ($0$ to $-4~\mathrm{eV}$) feeds widespread population of the conduction manifold up to $+4.9~\mathrm{eV}$. Furthermore, the marked difference between transient (green) and permanent (red) distributions at $+3.5~\mathrm{eV}$ and $+4.0~\mathrm{eV}$ highlights strong intra-pulse dynamical polarization prior to population relaxation into lower conduction states.

Comparison between the results obtained for 1T and 4T demonstrates that extended $\pi$-conjugation lowers the energy and field thresholds for populating higher virtual states. While 1T requires intense fields ($\eta \sim 10^{-1}$) to access virtual orbitals above 9~eV, 4T achieves broad, multi-orbital population across the conduction manifold under moderate driving ($\eta \sim 10^{-2}$). This sequential, many-particle redistribution provides the microscopic mechanism responsible for the sublinear saturation of $N_{\mathrm{ex}}$, the activation of excited-state absorption ($S_{\mathrm B} \to S_n$), and the emergence of high-harmonic spectral broadening.

\section{Conclusions}
\label{sec:conclusions}

In summary, we have investigated the ultrafast nonlinear optical response and laser-driven population dynamics in thiophene oligomers, taking 1T and 4T as representatives for odd- and even-numbered members, across different regimes of excitation. By inspecting energy uptake, the number of excited electrons, HHG spectra, and performing many-body population analysis, we have established a unified microscopic picture linking extended $\pi$-conjugation and molecular symmetry to intense-field performance.

Our findings demonstrate that the conjugated backbone from 1T to 4T drastically lowers the intensity threshold for nonlinear absorption, driving ground-state saturation at significantly weaker field strengths. Through state-resolved transient population analysis, we reveal that this saturation is accompanied by dynamic, field-induced channel switching, specifically, sequential population transfer from the pumped bright state into higher-lying electronic states, providing real-time dynamic evidence of ESA. 
The many-body population analysis scheme proposed and applied here shows that ground-state depletion and excited-state absorption are non-independent, sequential stages of a single field-driven pathway, $S_0\to S_{\mathrm B}\to S_n$, whose onset coincides with the crossover from the perturbative to the non-perturbative excitation regime. The pronounced sensitivity of $S_{\mathrm B}\to S_n$ channel activation to $\pi$-delocalization underscores the need to account for transient excited-state manifolds when designing organic chromophores for intense-field and attosecond spectroscopic applications.

Importantly, the field intensity range predicted for the non-perturbative crossover is experimentally compatible with pulsed excitations of molecular solids. In particular, a recent HHG experiment on pentacene single crystals reported efficient harmonic emission up to the 17th order under $4~\upmu$m, $103$-fs pump pulses at a peak intensity of $0.99~\mathrm{TW\,cm^{-2}}$, with the crystal remaining visibly intact and undamaged at this fluence~\cite{wiec+25natcomm}. Ultrafast pump-probe measurements on thin oligothiophene films, directly relevant to the systems studied here, have likewise resolved exciton population and relaxation dynamics on the femtosecond-to-picosecond timescale without reported sample degradation~\cite{vare+12prl}.  While detailed damage-threshold mapping for thiophene oligomers under the specific conditions simulated here remains a target for future experiments, our results indicate that the non-perturbative regime identified in our calculations is readily accessible in non-destructive laboratory setups.

\section*{Supplementary Material}

See the supplementary material for additional computational details, including:
\begin{itemize}
    \item the Thouless/CIS population-estimator derivation and excitation-rank decomposition;
    \item the full mapping between peak intensity and the dimensionless coupling parameter $\eta$;
    \item linear-response TDDFT benchmarks and Casida excitation assignments.
\end{itemize}

\begin{acknowledgments}
This work was funded by the German Research Foundation, Project numbers 524452181 (INPULS) and 398816777 (subproject A08). Computational resources were provided by the North-German Supercomputing Alliance (NHR), project nip00074.
\end{acknowledgments}

\section*{Author Declarations}

\subsection*{Conflict of Interest}
The authors have no conflicts to disclose.

\subsection*{Author Contributions}
\textbf{Mustapha Driouech:} Investigation, Formal analysis, Software, Data curation,
Visualization, Writing -- original draft.
\textbf{Caterina Cocchi:} Conceptualization, Funding acquisition, 
Supervision, Writing -- review \& editing.
\textbf{Michele Guerrini:} Conceptualization, Methodology, Software, Data
curation, Formal analysis, Validation, Visualization, Writing -- original draft,
Supervision.

\section*{Data Availability Statement}
The data supporting this study, including the curated Octopus input/output files,
projection data, processed population-analysis datasets, figure-ready tables, and
post-processing scripts, are openly available in Zenodo at \href{https://doi.org/10.5281/zenodo.22858413}{10.5281/zenodo.22858413}. 


\renewcommand{\bibsection}{\section*{References}}
\bibliography{references}

\end{document}


\maketitle
\tableofcontents
\newpage

\section{Many-body Population Framework and Operational Coupling-Regime Thresholds}
\label{app:population_estimators}

\subsection{Determinant overlaps and the Cramer's-rule identity}
\label{app:thouless_eval}

To track non-perturbative electronic dynamics, the time-evolving Kohn--Sham determinant $|\Phi(t)\rangle$ is projected onto field-free reference states. Let $\{\phi_p^0\}$ denote the set of field-free reference orbitals ($p=1,\ldots,N_{\mathrm{tot}}$, spanning occupied $i,j,\ldots$ and virtual $a,b,\ldots$ spaces), and let $\{\phi_n(t)\}$ be the $N_{\mathrm{occ}}$ time-propagated occupied orbitals. The overlap matrix $C(t)$ between the fixed, field-free reference Kohn--Sham orbitals $\{\phi_p^0\}$ ($p=1,\ldots,N_{\mathrm{tot}}$, running over occupied and virtual orbitals) and the $N_{\mathrm{occ}}$ propagated occupied orbitals $\{\phi_n(t)\}$ is
\begin{equation}
C_{pn}(t)=\langle\phi_p^0|\phi_n(t)\rangle,
\qquad
C(t)=
\begin{pmatrix}
A(t)\\
B(t)
\end{pmatrix}.
\end{equation}
Splitting the reference index into occupied ($i,j,\ldots$) and virtual ($a,b,\ldots$) blocks gives the $N_{\mathrm{occ}}\times N_{\mathrm{occ}}$ occupied--occupied block $A_{in}(t)=\langle\phi_i^0|\phi_n(t)\rangle$ and the $N_{\mathrm{virt}}\times N_{\mathrm{occ}}$ virtual--occupied block $B_{an}(t)=\langle\phi_a^0|\phi_n(t)\rangle$. The ground-state population is $P_{\mathrm{GS}}(t)=|\det A(t)|^2$, while the determinant expression for a configuration interaction single (CIS)/Casida state $I$ is
\begin{equation}
P_I(t)=
\left|
\sum_{ia}X_{ia}^{(I)*}
\det\bigl(A(t)|_{i\to a}\bigr)
\right|^2 .
\label{eq:SI_PI_manybody}
\end{equation}
$A(t)|_{i\to a}$ denotes the $N_{\mathrm{occ}}\times N_{\mathrm{occ}}$ matrix obtained from $A(t)$ by replacing its row for reference orbital $\phi_i^0$ with the corresponding row of $B(t)$ for reference orbital $\phi_a^0$; the remaining $N_{\mathrm{occ}}-1$ rows and all $N_{\mathrm{occ}}$ columns are unchanged, so $A(t)|_{i\to a}$ is a hybrid matrix, not $B(t)$ itself (which has $N_{\mathrm{virt}}$ rows in total). To evaluate Eq.~\eqref{eq:SI_PI_manybody}, we exploit the identity
\begin{equation}
\det\bigl(A(t)|_{i\to a}\bigr)=\det A(t)\cdot\bigl(B(t)A^{-1}(t)\bigr)_{ai},
\label{eq:cramer}
\end{equation}
which follows from Cramer's rule. Expanding $\det(A(t)|_{i\to a})$ by cofactors along its substituted row $i$, the remaining $N_{\mathrm{occ}}-1$ rows are identical to those of $A(t)$, so the cofactors along row $i$ equal the cofactors of $A(t)$ itself along that row:
\begin{equation}
\det\bigl(A(t)|_{i\to a}\bigr)=\sum_n B_{an}(t)\,C^A_{in}(t),
\end{equation}
where $C^A_{in}(t)$ is the $(i,n)$ cofactor of $A(t)$. Cramer's rule expresses the matrix inverse through the transposed cofactor (adjugate) matrix, $(A^{-1})_{ni}(t)=C^A_{in}(t)/\det A(t)$, i.e. $C^A_{in}(t)=\det A(t)\cdot(A^{-1})_{ni}(t)$; substituting back gives Eq.~\eqref{eq:cramer}. This reduces the evaluation of $P_I(t)$ to a single $O(N_{\mathrm{occ}}^3)$ matrix solve per time step, rather than $O(N_{\mathrm{occ}}N_{\mathrm{virt}})$ separate determinants.

The quantity $Z(t)\equiv B(t)A^{-1}(t)$ appearing in Eq.~\eqref{eq:cramer} is the amplitude of an equivalent, exact exponential (Thouless) representation of the propagated determinant relative to the reference $|\Phi_0\rangle$. Because $|\Phi(t)\rangle$ is itself a single Slater determinant rather than a genuinely correlated many-body state, this single-excitation exponential parametrization is not a truncated CIS or coupled-cluster ansatz but an exact orbital-rotation identity that closes after one term~\cite{thou60np,ring-schu80,helg-jorg-olse00}: provided $A(t)$ is nonsingular,
\begin{equation}
|\Phi(t)\rangle=\det A(t)\,\exp\!\left[\sum_{ia}Z_{ai}(t)c_a^\dagger c_i\right]|\Phi_0\rangle,
\qquad
Z(t)=B(t)A^{-1}(t).
\label{eq:thouless_exp}
\end{equation}

Summing the Cramer's-rule identity above over $(i,a)$ weighted by $X_{ia}^{(I)*}$ and comparing with Eq.~\eqref{eq:SI_PI_manybody} gives
\begin{align}
\langle\Phi_I|\Phi(t)\rangle&=\sum_{ia}X_{ia}^{(I)*}\det\bigl(A(t)|_{i\to a}\bigr)\notag\\
&=\det A(t)\sum_{ia}X_{ia}^{(I)*}Z_{ai}(t)\notag\\
&\equiv\det A(t)\,\mathcal{Z}_I(t),
\end{align}
so the Casida-projected Thouless amplitude
\begin{equation}
\mathcal{Z}_I(t)=\sum_{ia}X_{ia}^{(I)*}Z_{ai}(t)
\end{equation}
is exactly the excited-state overlap $\langle\Phi_I|\Phi(t)\rangle$, normalized by the (complex) ground-state overlap amplitude $\det A(t)$ -- not an independently defined quantity. Squaring and using $P_0(t)=|\det A(t)|^2$, Eq.~\eqref{eq:SI_PI_manybody} becomes the exact reformulation
\begin{equation}
P_I(t)=P_0(t)\,|\mathcal{Z}_I(t)|^2 \equiv P_I^{\mathrm{Th}}(t),
\label{eq:PThou}
\end{equation}
which is the form we evaluate throughout this work.

\subsection{The CIS/1RDM population estimator as the weak-field limit of the Thouless representation}
\label{app:coh_incoh}

To connect the determinant-overlap framework with conventional density-matrix analyses, we introduce the instantaneous one-particle reduced density matrix (1RDM) in the reference basis, $\rho_{pq}(t)=\langle c_q^\dagger c_p\rangle_t=\langle\Phi(t)|c_q^\dagger c_p|\Phi(t)\rangle=[C(t)C^\dagger(t)]_{pq}$; in block form, $\rho_{oo}(t)=A(t)A^\dagger(t)$, $\rho_{vv}(t)=B(t)B^\dagger(t)$, and the particle-hole block $\rho_{ai}(t)=[B(t)A^\dagger(t)]_{ai}$. The standard CIS excited-state population estimator is obtained by contracting the particle--hole density-matrix kernel with the CIS/Casida coefficients
\begin{align}
P_I^{\mathrm{CIS}}(t)&=\sum_{ia,jb}X_{ia}^{(I)*}X_{jb}^{(I)}K_{ia,jb}(t),\notag\\
K_{ia,jb}(t)&=\langle c_a^\dagger c_i c_j^\dagger c_b\rangle_t=\langle\Phi(t)|c_a^\dagger c_i c_j^\dagger c_b|\Phi(t)\rangle,
\end{align}
where the particle--hole kernel is built directly from $\rho_{ai}(t)$, \emph{without} the $A^{-1}(t)$ renormalization entering the Thouless amplitude $Z(t)=B(t)A^{-1}(t)$. Although $K_{ia,jb}(t)$ carries four operator indices and looks like a two-particle (2RDM) quantity, for a single TDKS determinant it is not: as shown explicitly below, Wick's theorem reduces it exactly to a product of one-particle (1RDM) elements, with no independent two-body content. We therefore refer to $P_I^{\mathrm{CIS}}(t)$ as the CIS/1RDM estimator throughout, since it is entirely a functional of $\rho_{pq}(t)$.

Since $|\Phi(t)\rangle$ is a single Slater determinant, Wick's theorem~\cite{lowd55pr} reduces the four-operator expectation value $K_{ia,jb}(t)$ to a sum over all pairwise contractions of $(c_a^\dagger,c_i,c_j^\dagger,c_b)$ into the nonzero one-body contractions $\langle c_p^\dagger c_q\rangle_t=\rho_{qp}(t)$ and $\langle c_pc_q^\dagger\rangle_t=\delta_{pq}-\rho_{pq}(t)$ (contractions between two creation or two annihilation operators vanish, since $|\Phi(t)\rangle$ has fixed particle number). The two nonvanishing pairings, $(c_a^\dagger c_i)(c_j^\dagger c_b)\to\rho_{ia}(t)\rho_{bj}(t)$ and $(c_a^\dagger c_b)(c_ic_j^\dagger)\to\rho_{ba}(t)[\delta_{ij}-\rho_{ij}(t)]$, both enter with sign $+1$, giving
\begin{equation}
K_{ia,jb}(t)=\rho_{ia}(t)\rho_{bj}(t)+\rho_{ba}(t)\bigl[\delta_{ij}-\rho_{ij}(t)\bigr].
\label{eq:Wick_kernel}
\end{equation}
Every term on the right-hand side of Eq.~\eqref{eq:Wick_kernel} is a product of two 1RDM elements: this is the precise sense in which a formally two-particle kernel carries no independent 2RDM information for a single determinant, confirming that $P_I^{\mathrm{CIS}}(t)$ is fully determined by $\rho_{pq}(t)$ alone. Inserting Eq.~\eqref{eq:Wick_kernel} into $P_I^{\mathrm{CIS}}(t)$, the first term factorizes exactly, using the Hermiticity $\rho_{bj}=\rho_{jb}^*$,
\begin{equation}
\sum_{ia,jb}X_{ia}^{(I)*}X_{jb}^{(I)}\rho_{ia}(t)\rho_{bj}(t)=\Bigl|\sum_{ia}X_{ia}^{(I)*}\rho_{ia}(t)\Bigr|^2\equiv P_I^{\mathrm{coh}}(t),
\end{equation}
the \emph{coherent} particle-hole contribution. The diagonal ($i=j$, $a=b$) restriction of the second term gives an independent, additive \emph{incoherent} contribution,
\begin{align}
P_I^{\mathrm{diag}}(t) &= \sum_{ia}|X_{ia}^{(I)}|^2\,n_a(t)\,h_i(t), \nonumber \\ & n_a(t)=\rho_{aa}(t),\ \ h_i(t)=1-\rho_{ii}(t),
\end{align}
i.e. the population state $I$ would have if the virtual occupation $n_a(t)$ and the hole population $h_i(t)$ were statistically independent. The remaining, off-diagonal ($i\ne j$ and/or $a\ne b$) part of the second term in eq.~\eqref{eq:Wick_kernel} contributes further intraband occupied and unoccupied coherence corrections that we do not track separately here.

Because $\rho(t)$ is all that enters $P_I^{\mathrm{CIS}}(t)$ [Eq.~\eqref{eq:Wick_kernel}], the CIS/1RDM estimator and the exact many-body estimator $P_I^{\mathrm{Th}}(t)$ [Eq.~\eqref{eq:PThou}] differ only through the single-particle quantity each one uses in place of the true occupied-to-virtual admixture: $\rho_{ai}(t)=[B(t)A^\dagger(t)]_{ai}$ for CIS/1RDM, versus the properly renormalized Thouless amplitude $Z_{ai}(t)=[B(t)A^{-1}(t)]_{ai}$ for $P_I^{\mathrm{Th}}(t)$. In the perturbative, weak-field regime the occupied--occupied block approaches the identity, $A(t)\to I$, so $Z(t)\to B(t)$ and $\rho_{ai}(t)\to B_{ai}(t)$ in the same limit: $Z_{ai}(t)$ and $\rho_{ai}(t)$ coincide to leading order in the field amplitude, and, substituting $P_0(t)\to1$ and $\mathcal{Z}_I(t)\to\sum_{ia}X_{ia}^{(I)*}B_{ai}(t)$ into Eq.~\eqref{eq:PThou}, $P_I^{\mathrm{coh}}(t)$ above is recovered as exactly the leading-order term of $P_I^{\mathrm{Th}}(t)$ in this same limit. This is the precise sense in which the CIS/1RDM estimator is a weak-field special case of the many-body framework of Sec.~II.B, collapsing the full Thouless formulation onto its bare-1RDM leading order. Beyond the weak-field regime this collapse is no longer innocuous: $\rho(t)$ lacks the $A^{-1}(t)$ renormalization by ground-state depletion built into $Z(t)$, and using the bare $\rho(t)$ at all orders is precisely what introduces the extra, independent term $P_I^{\mathrm{diag}}(t)$, absent from the exact many-body theory. Consequently $\sum_I P_I^{\mathrm{CIS}}(t)$ can exceed unity once $P_I^{\mathrm{diag}}(t)$ is non-negligible, while the exact sum rule [Eq.~\eqref{eq:SI_Pge2_sumrule} below] holds at every intensity.

Since the propagated orbitals remain orthonormal, $A^\dagger(t)A(t)+B^\dagger(t)B(t)=I$, so $A^\dagger(t)A(t)=I-\Delta(t)$ exactly, with $\Delta(t)\equiv B^\dagger(t)B(t)$. Using $P_0(t)=|\det A(t)|^2=\det[A^\dagger(t)A(t)]$,
\begin{equation}
P_0(t)=\det\bigl(I-\Delta(t)\bigr)=1-\operatorname{Tr}\Delta(t)+O(\epsilon^4),
\label{eq:P0split}
\end{equation}
where $\epsilon\propto E_0$ is the field-amplitude bookkeeping parameter and, in the spin-orbital convention used here, $\operatorname{Tr}\Delta(t)=\sum_{ia}|B_{ai}(t)|^2$ is exactly the single-particle excited-electron count $N_{\mathrm{ex}}(t)$ used in the main text. We use this identity below (Sec.~\ref{app:rank}) as an independent check of the excitation-rank sum rule.

\subsection{Excitation-rank decomposition and non-perturbativity descriptor}
\label{app:rank}
\label{app:nonpert}

The Thouless representation also gives the exact rank content of $|\Phi(t)\rangle$. Expanding the exponential in Eq.~\eqref{eq:thouless_exp} as a finite sum,
\begin{equation}
|\Phi(t)\rangle=\det A(t)\sum_{r=0}^{N_{\mathrm{occ}}}\frac{1}{r!}\hat T_1(t)^r|\Phi_0\rangle,
\label{eq:rank_expansion}
\end{equation}
truncating exactly at $r=N_{\mathrm{occ}}$ since $\hat T_1(t)^r|\Phi_0\rangle=0$ once $r$ exceeds the number of occupied indices available to excite. Each term $\hat T_1(t)^r|\Phi_0\rangle$ is, up to a combinatorial prefactor, a linear combination of $r$-fold particle-hole excited determinants relative to $\Phi_0$: it lies entirely within the rank-$r$ sector of Fock space, spanned by determinants differing from $\Phi_0$ in exactly $r$ orbitals. Determinants built from an orthonormal single-particle basis with different occupied-orbital sets are automatically orthogonal, so the rank-$r$ sectors for $r=0,1,\ldots,N_{\mathrm{occ}}$ are mutually orthogonal, and every cross term in $\langle\Phi(t)|\Phi(t)\rangle$ vanishes. Since $|\Phi(t)\rangle$ is a normalized TDKS state at every time,
\begin{equation}
1=\langle\Phi(t)|\Phi(t)\rangle=|\det A(t)|^2\sum_{r=0}^{N_{\mathrm{occ}}}\left\|\frac{1}{r!}\hat T_1(t)^r|\Phi_0\rangle\right\|^2 .
\end{equation}
Defining the rank-$r$ population as $P_r(t)\equiv P_0(t)\,\|\hat T_1(t)^r|\Phi_0\rangle/r!\|^2$, the exact sum rule $\sum_{r=0}^{N_{\mathrm{occ}}}P_r(t)=1$ therefore follows directly from normalization of $|\Phi(t)\rangle$ together with the mutual orthogonality of distinct excitation-rank sectors -- with no further input, and independently of the field strength or of any weak-field approximation.

A standard identity for antisymmetrized powers of a linear map~\cite{zang+20prl} evaluates each rank-$r$ norm explicitly in terms of the eigenvalues $\{\lambda_m(t)\}$ of $Z(t)Z^\dagger(t)$:
\begin{equation}
\left\|\frac{1}{r!}\hat T_1(t)^r|\Phi_0\rangle\right\|^2
=
e_r[\{\lambda_m(t)\}],
\end{equation}
where $e_r$ is the elementary symmetric polynomial of order $r$,
\begin{equation}
e_r[\{\lambda_m\}]
=
\sum_{1\le m_1<\cdots<m_r\le M}
\lambda_{m_1}\lambda_{m_2}\cdots\lambda_{m_r},
\qquad
e_0=1 .
\end{equation}
Here $M=\mathrm{rank}[Z(t)Z^\dagger(t)]\le N_{\mathrm{occ}}$. Thus $e_1=\sum_m\lambda_m$, $e_2=\sum_{m<n}\lambda_m\lambda_n$, and higher orders are built analogously from products of distinct eigenvalues. These higher-rank terms arise from antisymmetrized products of $Z_{ai}(t)$ alone, not independent correlation amplitudes -- unlike a general coupled-cluster ansatz, a single determinant carries no independent doubles amplitude. The rank-$r$ population is therefore $P_r(t)=P_0(t)\,e_r[\{\lambda_m(t)\}]$. As an explicit algebraic check on the normalization argument above, the generating function of the elementary symmetric polynomials, $\sum_{r=0}^{N_{\mathrm{occ}}}e_r[\{\lambda_m\}]x^r=\prod_m(1+\lambda_mx)$, evaluated at $x=1$ gives $\sum_rP_r(t)=P_0(t)\prod_m\bigl(1+\lambda_m(t)\bigr)=P_0(t)\det\bigl(I+Z(t)Z^\dagger(t)\bigr)$. Using $Z=BA^{-1}$ and Sylvester's determinant identity $\det(I_{N_{\mathrm{virt}}}+XY)=\det(I_{N_{\mathrm{occ}}}+YX)$ for rectangular $X,Y$,
\begin{align}
\det\bigl(I+ZZ^\dagger\bigr)&=\det\bigl(I+B(A^\dagger A)^{-1}B^\dagger\bigr)\notag\\
&=\det\bigl(I+(A^\dagger A)^{-1}B^\dagger B\bigr)\notag\\
&=\det\bigl(I+(I-\Delta)^{-1}\Delta\bigr)\notag\\
&=\det\bigl[(I-\Delta)^{-1}\bigr]=\frac{1}{P_0(t)},
\end{align}
using $A^\dagger(t)A(t)=I-\Delta(t)$, $B^\dagger(t)B(t)=\Delta(t)$ [Sec.~\ref{app:coh_incoh}], and $P_0(t)=\det(I-\Delta(t))$ [Eq.~\eqref{eq:P0split}]. Hence $P_0(t)\det(I+ZZ^\dagger(t))=1$ exactly, independently confirming the sum rule already established by normalization:
\begin{equation}
P_0(t)+\underbrace{\textstyle P_0(t)\sum_m\lambda_m(t)}_{P_1^{\mathrm{tot}}(t)}+\,P_{\geq2}(t)=1,
\label{eq:SI_Pge2_sumrule}
\end{equation}
which defines the double-and-higher excitation weight $P_{\geq2}(t)$; only the tracked Casida states enter $P_1^{\mathrm{Casida}}(t)=\sum_I P_I^{\mathrm{Th}}(t)\le P_1^{\mathrm{tot}}(t)$.

The residual cumulative probability function $P_{\geq2}(t)$ quantifies the fraction of the propagated TDKS determinant that lies outside the ground-state plus single electron--hole excitation sectors, i.e., the weight carried by double-and-higher electron--hole excitation orders relative to the field-free reference determinant. It therefore distinguishes a response dominated by ground-state depletion and single electron--hole excitations from one in which higher excitation orders acquire appreciable weight. In this sense, $P_{\geq2}(t)$ is a diagnostic of when the singles-only (CIS/1RDM) description above breaks down: $P_{\geq2}(t)\approx0$ is the regime in which that description remains valid, while a non-negligible $P_{\geq2}(t)$ signals double-and-higher excitation orders generated by nonlinear antisymmetrized products of the Thouless singles, with no singles-only counterpart. We use its peak value along a trajectory, $P_{\geq2}^{\max}\equiv\max_t P_{\geq2}(t)$, as a non-perturbativity descriptor.

\subsection{Intensity--$\eta$ Mapping and Operational coupling-regime thresholds}
\label{sec:SI_eta_mapping}

The conversion between peak laser intensity $I$ ($\text{GW\,cm}^{-2}$) and electric field amplitude $E_0$ (atomic units) is evaluated via:
\begin{equation}
E_0[\mathrm{a.u.}] = \sqrt{\frac{I[\mathrm{GW\,cm^{-2}}]}{3.50944506\times10^{7}}}.
\end{equation}
The dimensionless dipolar coupling strength $\eta = |E_0\mu_\parallel| / \hbar\omega$ is evaluated in a consistent atomic-unit convention using the bright-state excitation parameters specified in the main text ($\hbar\omega=5.5$~eV, $\mu_\parallel=0.384$~\AA{} for 1T; $\hbar\omega=2.5$~eV, $\mu_\parallel=2.135$~\AA{} for 4T). 

Table~\ref{tab:eta_mapping} provides the point-by-point numerical mapping between peak pump intensity, $\eta$, post-pulse energy uptake $\Delta E(t_f)$, excited electron count $N_{\mathrm{ex}}(t_f)$, peak bright-state population $P_{S_{\mathrm B}}^{\max}$, and peak non-perturbative weight leakage $P_{\geq2}^{\max}$ (evaluated over $t\le20$~fs for 1T and $t\le27$~fs for 4T). 

Table~\ref{tab:local_alpha} lists the local power-law scaling exponents $\alpha_j^{(Q)}$ extracted across intensity intervals $(I_j, I_{j+1})$ via:
\begin{equation}
\alpha_j^{(Q)} = \frac{\log\!\left[Q(I_{j+1})/Q(I_j)\right]}{\log\!\left(I_{j+1}/I_j\right)},
\label{eq:SI_local_alpha}
\end{equation}
where $Q \in \{\Delta E(t_f), N_{\mathrm{ex}}(t_f)\}$. In the weak-field one-photon regime, $\alpha_j^{(Q)}\approx1$; deviations from unity indicate population saturation ($\alpha < 1$) or higher-order multi-photon processes ($\alpha > 1$).

\begin{table}[p]
\centering
\caption{Intensity--$\eta$ map and response diagnostics for 1T and 4T. $\Delta E(t_f)$ is the post-pulse energy uptake and $N_{\mathrm{ex}}(t_f)$ is the post-pulse number of excited electrons. $P_{S_{\mathrm B}}^{\max}$ is the maximum pumped bright-state population, while $P_{\geq2}^{\max}$ is the maximum leakage outside the ground-state plus single-excitation sector.}
\label{tab:eta_mapping}
\small
\setlength{\tabcolsep}{0pt}
\begin{tabular*}{\textwidth}{@{\extracolsep{\fill}}lccccc@{}}
\hline \hline
Pump intensity & $\eta$ & $\Delta E(t_f)$ & $N_{\mathrm{ex}}(t_f)$ & $P_{S_{\mathrm B}}^{\max}$ & $P_{\geq2}^{\max}$ \\
(GW cm$^{-2}$) & & (eV) & & & \\
\hline
\multicolumn{6}{l}{\textbf{1T}} \\
10 & $1.917\times10^{-3}$ & 0.017 & 0.003 & $1.354\times10^{-3}$ & $1.391\times10^{-5}$ \\
20 & $2.711\times10^{-3}$ & 0.034 & 0.005 & $2.704\times10^{-3}$ & $2.413\times10^{-5}$ \\
50 & $4.286\times10^{-3}$ & 0.085 & 0.013 & $6.730\times10^{-3}$ & $5.648\times10^{-5}$ \\
100 & $6.062\times10^{-3}$ & 0.171 & 0.027 & $1.336\times10^{-2}$ & $1.165\times10^{-4}$ \\
200 & $8.573\times10^{-3}$ & 0.342 & 0.053 & $2.636\times10^{-2}$ & $2.609\times10^{-4}$ \\
250 & $9.585\times10^{-3}$ & 0.428 & 0.066 & $3.274\times10^{-2}$ & $3.491\times10^{-4}$ \\
500 & $1.355\times10^{-2}$ & 0.857 & 0.142 & $6.339\times10^{-2}$ & $9.228\times10^{-4}$ \\
1000 & $1.917\times10^{-2}$ & 1.709 & 0.327 & $1.187\times10^{-1}$ & $2.853\times10^{-3}$ \\
2000 & $2.711\times10^{-2}$ & 3.344 & 0.508 & $2.039\times10^{-1}$ & $9.547\times10^{-3}$ \\
5000 & $4.286\times10^{-2}$ & 7.734 & 0.871 & $3.006\times10^{-1}$ & $3.998\times10^{-2}$ \\
10000 & $6.062\times10^{-2}$ & 12.878 & 1.284 & $3.164\times10^{-1}$ & $1.126\times10^{-1}$ \\
\hline
\multicolumn{6}{l}{\textbf{4T}} \\
10 & $2.344\times10^{-2}$ & 0.951 & 0.382 & $1.611\times10^{-1}$ & $2.072\times10^{-3}$ \\
20 & $3.315\times10^{-2}$ & 1.743 & 0.715 & $2.510\times10^{-1}$ & $8.489\times10^{-3}$ \\
30 & $4.060\times10^{-2}$ & 2.376 & 0.888 & $2.982\times10^{-1}$ & $2.036\times10^{-2}$ \\
40 & $4.688\times10^{-2}$ & 2.892 & 0.986 & $3.277\times10^{-1}$ & $3.411\times10^{-2}$ \\
50 & $5.242\times10^{-2}$ & 3.330 & 1.054 & $3.486\times10^{-1}$ & $4.751\times10^{-2}$ \\
60 & $5.742\times10^{-2}$ & 3.713 & 1.105 & $3.646\times10^{-1}$ & $6.004\times10^{-2}$ \\
70 & $6.202\times10^{-2}$ & 4.058 & 1.142 & $3.774\times10^{-1}$ & $7.263\times10^{-2}$ \\
80 & $6.630\times10^{-2}$ & 4.373 & 1.170 & $3.881\times10^{-1}$ & $8.491\times10^{-2}$ \\
90 & $7.032\times10^{-2}$ & 4.666 & 1.191 & $3.972\times10^{-1}$ & $9.600\times10^{-2}$ \\
100 & $7.413\times10^{-2}$ & 4.941 & 1.208 & $4.050\times10^{-1}$ & $1.069\times10^{-1}$ \\
1000 & $2.344\times10^{-1}$ & 19.613 & 2.493 & $5.267\times10^{-1}$ & $5.344\times10^{-1}$ \\
2000 & $3.315\times10^{-1}$ & 33.915 & 3.516 & $5.654\times10^{-1}$ & $7.317\times10^{-1}$ \\
2500 & $3.706\times10^{-1}$ & 40.908 & 3.963 & $5.638\times10^{-1}$ & $7.930\times10^{-1}$ \\
4000 & $4.688\times10^{-1}$ & 60.940 & 5.288 & $5.716\times10^{-1}$ & $9.189\times10^{-1}$ \\
5000 & $5.242\times10^{-1}$ & 70.101 & 5.904 & $5.800\times10^{-1}$ & $9.505\times10^{-1}$ \\
6000 & $5.742\times10^{-1}$ & 86.368 & 6.536 & $6.023\times10^{-1}$ & $9.631\times10^{-1}$ \\
7000 & $6.202\times10^{-1}$ & 90.125 & 6.304 & $5.830\times10^{-1}$ & $9.675\times10^{-1}$ \\
8000 & $6.630\times10^{-1}$ & 108.424 & 7.411 & $5.856\times10^{-1}$ & $9.809\times10^{-1}$ \\
9000 & $7.032\times10^{-1}$ & 120.592 & 7.732 & $5.901\times10^{-1}$ & $9.866\times10^{-1}$ \\
10000 & $7.413\times10^{-1}$ & 133.654 & 8.187 & $5.939\times10^{-1}$ & $9.887\times10^{-1}$ \\
\hline \hline
\end{tabular*}
\end{table}

\begin{table}[p]
\centering
\caption{Intensity and $\eta$ dependent power--law scaling exponents for the post-pulse observables used in
Fig.~1 of the main text, calculated from Eq.~\eqref{eq:SI_local_alpha}.}
\label{tab:local_alpha}
\scriptsize
\setlength{\tabcolsep}{0pt}
\begin{tabular*}{\textwidth}{@{\extracolsep{\fill}}lccc@{}}
\hline \hline
Pump interval & $\eta$ interval & $\alpha_j^{(\Delta E)}$ & $\alpha_j^{(N_{\mathrm{ex}})}$ \\
(GW cm$^{-2}$) & & & \\
\hline
\multicolumn{4}{l}{\textbf{1T}} \\
10--20 & $1.92\times10^{-3}$--$2.71\times10^{-3}$ & 0.99 & 0.99 \\
20--50 & $2.71\times10^{-3}$--$4.29\times10^{-3}$ & 1.00 & 0.99 \\
50--100 & $4.29\times10^{-3}$--$6.06\times10^{-3}$ & 1.00 & 0.98 \\
100--200 & $6.06\times10^{-3}$--$8.57\times10^{-3}$ & 1.00 & 0.99 \\
200--250 & $8.57\times10^{-3}$--$9.58\times10^{-3}$ & 1.00 & 1.02 \\
250--500 & $9.58\times10^{-3}$--$1.36\times10^{-2}$ & 1.00 & 1.10 \\
500--1000 & $1.36\times10^{-2}$--$1.92\times10^{-2}$ & 1.00 & 1.20 \\
1000--2000 & $1.92\times10^{-2}$--$2.71\times10^{-2}$ & 0.97 & 0.64 \\
2000--5000 & $2.71\times10^{-2}$--$4.29\times10^{-2}$ & 0.92 & 0.59 \\
5000--10000 & $4.29\times10^{-2}$--$6.06\times10^{-2}$ & 0.74 & 0.56 \\
\hline
\multicolumn{4}{l}{\textbf{4T}} \\
10--20 & $2.34\times10^{-2}$--$3.32\times10^{-2}$ & 0.87 & 0.90 \\
20--30 & $3.32\times10^{-2}$--$4.06\times10^{-2}$ & 0.76 & 0.54 \\
30--40 & $4.06\times10^{-2}$--$4.69\times10^{-2}$ & 0.68 & 0.36 \\
40--50 & $4.69\times10^{-2}$--$5.24\times10^{-2}$ & 0.63 & 0.30 \\
50--60 & $5.24\times10^{-2}$--$5.74\times10^{-2}$ & 0.60 & 0.26 \\
60--70 & $5.74\times10^{-2}$--$6.20\times10^{-2}$ & 0.58 & 0.22 \\
70--80 & $6.20\times10^{-2}$--$6.63\times10^{-2}$ & 0.56 & 0.18 \\
80--90 & $6.63\times10^{-2}$--$7.03\times10^{-2}$ & 0.55 & 0.15 \\
90--100 & $7.03\times10^{-2}$--$7.41\times10^{-2}$ & 0.54 & 0.13 \\
100--1000 & $7.41\times10^{-2}$--$2.34\times10^{-1}$ & 0.60 & 0.31 \\
1000--2000 & $2.34\times10^{-1}$--$3.32\times10^{-1}$ & 0.79 & 0.50 \\
2000--2500 & $3.32\times10^{-1}$--$3.71\times10^{-1}$ & 0.84 & 0.54 \\
2500--4000 & $3.71\times10^{-1}$--$4.69\times10^{-1}$ & 0.85 & 0.61 \\
4000--5000 & $4.69\times10^{-1}$--$5.24\times10^{-1}$ & 0.63 & 0.49 \\
5000--6000 & $5.24\times10^{-1}$--$5.74\times10^{-1}$ & 1.14 & 0.56 \\
6000--7000 & $5.74\times10^{-1}$--$6.20\times10^{-1}$ & 0.28 & -0.23 \\
7000--8000 & $6.20\times10^{-1}$--$6.63\times10^{-1}$ & 1.38 & 1.21 \\
8000--9000 & $6.63\times10^{-1}$--$7.03\times10^{-1}$ & 0.90 & 0.36 \\
9000--10000 & $7.03\times10^{-1}$--$7.41\times10^{-1}$ & 0.98 & 0.54 \\
\hline \hline
\end{tabular*}
\end{table}

\clearpage

\newpage
\section{Linear Response Analysis} \label{Casida}

\begin{figure}[H]
  \centering
  \includegraphics[width=0.9\linewidth]{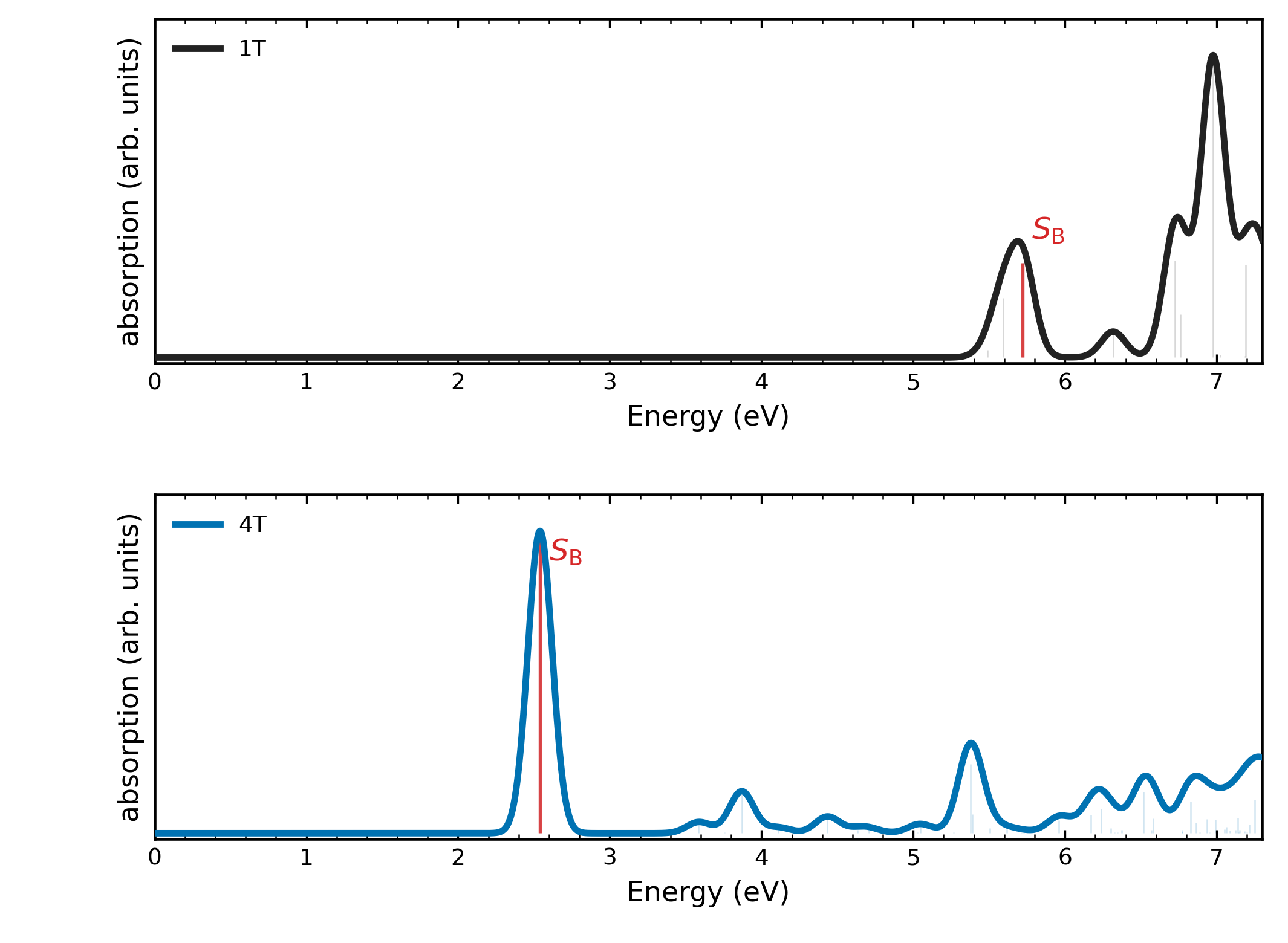}
  \caption{Casida absorption spectra of 1T and 4T from \textsc{Octopus} using the LDA functional. The red sticks highlight the pumped bright state $S_{\mathrm B}$ primarily considered in our analysis.}
  \label{fig:casida_spectra_si}
\end{figure}

In Fig.~\ref{fig:casida_spectra_si}, we display the first five excitations for 1T and 4T computed with \textsc{Octopus} using the Casida scheme for linear response (LDA functional). We checked that the first excitation predicted by Casida in the spectrum of 1T is a spurious effect of the local exchange-correlation functional and of the numerical grids implemented in \textsc{Octopus} (compare Tables~\ref{tab:excitation_1T_4T_gaussian} and \ref{tab:excitation_1T_4T_cam_b3lyp}). Due to its extremely weak oscillator strength, we can exclude any influence of this feature on the results of this work. 

\begin{table}[p]
\centering
\caption{Excitation energies, oscillator strengths ($f$), polarization directions, and molecular-orbital compositions of the first five excitations of 1T and 4T obtained from the linear-response Casida scheme implemented in \textsc{Octopus} using the LDA functional. Only orbital contributions larger than 10\% are included. H and L denote the HOMO and LUMO, respectively. The polarization direction is assigned according to the Cartesian component ($x$, $y$, or $z$) with the largest absolute transition-dipole moment.}
\label{tab:excitation_1T_4T}
\scriptsize
\setlength{\tabcolsep}{0pt}
\begin{tabular*}{\textwidth}{@{\extracolsep{\fill}}lcccl@{}}
\hline \hline
Excitation & Energy (eV) & $f$ & Pol. & Composition (weight in \%) \\
\hline
\multicolumn{5}{l}{\textbf{1T}} \\
1 & 5.49 & 0.002 & $ \langle \hat{z} \rangle $ & H $\to$ L+1 (99\%) \\
2 & 5.59 & 0.013 & $ \langle \hat{x}  \rangle $ & H-1 $\to$ L (75\%); H $\to$ L+3 (22\%) \\
3 & 5.65 & 0.000 & $ \langle \hat{z}  \rangle $ & H $\to$ L+2 (99\%) \\
4 & 5.72 & 0.021 & $ \langle \hat{y}  \rangle $ & H $\to$ L (93\%) \\
5 & 5.84 & 0.000 & $ \langle \hat{z}  \rangle $ & H-1 $\to$ L+1 (99\%) \\
\hline
\multicolumn{5}{l}{\textbf{4T}} \\
1 & 2.54 & 0.283 & $ \langle \hat{x}  \rangle $  & H $\to$ L (96\%) \\
2 & 2.81 & 0.000 & $ \langle \hat{x}  \rangle $   & H-1 $\to$ L (36\%); H $\to$ L+1 (62\%) \\
3 & 3.58 & 0.000 &  $ \langle \hat{z}  \rangle $   & H-1 $\to$ L (52\%); H $\to$ L+1 (24\%); H-2 $\to$ L (15\%) \\
4 & 3.59 & 0.010 &  $ \langle \hat{x}  \rangle $  & H-1 $\to$ L+1 (13\%); H $\to$ L+2 (71\%) \\
5 & 3.75 & 0.000 & $ \langle \hat{x}  \rangle $  & H-4 $\to$ L (19\%); H-2 $\to$ L (62\%) \\
\hline \hline
\end{tabular*}
\end{table}


In Table~\ref{tab:excitation_1T_4T_gaussian}, we report the energy, oscillator strength, and composition of the first 5 excitations computed from linear-response TDDFT using Gaussian16~\cite{g16} in the adiabatic LDA. 
Finally, in Table~\ref{tab:benchmark_summary}, we compare the energy of the first bright excitation computed for all molecules using LDA (RT-TDDFT and linear response with \textsc{Octopus} using the Perdew-Zunger functional~\cite{perd-zung81pr} and Gaussian16)  with experimental values recorded in solution at room temperature~\cite{beck+96jpch}. 

\begin{table}[p]
\centering
\caption{Excitation energies, oscillator strengths ($f$), and molecular-orbital compositions of the first five excitations of 1T and 4T obtained from TDDFT calculations using the LDA functional in Gaussian16. Only orbital contributions larger than 10\% are included. H and L denote the HOMO and LUMO, respectively.}
\label{tab:excitation_1T_4T_gaussian}
\scriptsize
\setlength{\tabcolsep}{0pt}
\begin{tabular*}{\textwidth}{@{\extracolsep{\fill}}cccl@{}}
\hline \hline
Excitation & Energy (eV) & $f$ & Composition (weight in \%) \\
\hline
\multicolumn{4}{l}{\textbf{1T}} \\
1 & 5.89 & 0.06 & H--1 $\rightarrow$ L (41\%) \\
2 & 5.93 & 0.00 & H $\rightarrow$ L+1 (50\%) \\
3 & 5.95 & 0.09 & H $\rightarrow$ L (48\%) \\
4 & 6.35 & 0.00 & H--1 $\rightarrow$ L+1 (50\%) \\
5 & 7.51 & 0.00 & H--2 $\rightarrow$ L (49\%) \\
\hline
\multicolumn{4}{l}{\textbf{4T}} \\
1 & 2.96 & 2.05 & H--1 $\rightarrow$ L+1 (21\%); H $\rightarrow$ L (66\%) \\
2 & 3.67 & 0.00 & H--1 $\rightarrow$ L (45\%); H $\rightarrow$ L+1 (51\%) \\
3 & 4.28 & 0.00 & H--1 $\rightarrow$ L (49\%); H $\rightarrow$ L+1 (43\%) \\
4 & 4.31 & 0.15 & H--1 $\rightarrow$ L+1 (41\%); H $\rightarrow$ L+2 (44\%) \\
5 & 4.86 & 0.00 & H--2 $\rightarrow$ L (47\%); H $\rightarrow$ L+2 (42\%) \\
\hline \hline
\end{tabular*}
\end{table}

\begin{table}[p]
\centering
\caption{Excitation energies, oscillator strengths ($f$), and molecular-orbital compositions of the first five excitations of 1T and 4T obtained from TDDFT calculations using the CAM-B3LYP functional in Gaussian16. Only orbital contributions larger than 10\% are included. H and L denote the HOMO and LUMO, respectively.}
\label{tab:excitation_1T_4T_cam_b3lyp}
\scriptsize
\setlength{\tabcolsep}{0pt}
\begin{tabular*}{\textwidth}{@{\extracolsep{\fill}}cccl@{}}
\hline \hline
Excitation & Energy (eV) & $f$ & Composition (weight in \%) \\
\hline
\multicolumn{4}{l}{\textbf{1T}} \\
1 & 6.10 & 0.09 & H $\rightarrow$ L (48\%) \\
2 & 6.19 & 0.10 & H--1 $\rightarrow$ L (44\%) \\
3 & 6.54 & 0.00 & H $\rightarrow$ L+1 (49\%) \\
4 & 6.56 & 0.00 & H--1 $\rightarrow$ L+1 (49\%) \\
5 & 8.07 & 0.00 & H--2 $\rightarrow$ L (48\%) \\
\hline
\multicolumn{4}{l}{\textbf{4T}} \\
1 & 3.32 & 1.21 & H $\rightarrow$ L (48\%) \\
2 & 4.33 & 0.00 & H--1 $\rightarrow$ L (16\%); H $\rightarrow$ L+1 (30\%) \\
3 & 4.79 & 0.00 & H--1 $\rightarrow$ L (30\%); H $\rightarrow$ L+1 (14\%) \\
4 & 5.06 & 0.03 & H--5 $\rightarrow$ L (16\%); H $\rightarrow$ L+2 (14\%) \\
5 & 5.11 & 0.00 & H--2 $\rightarrow$ L (35\%) \\
\hline \hline
\end{tabular*}
\end{table}

\begin{table}[H]
\centering
\caption{First excitation energy (in eV) of the oligothiophenes considered in this work computed from RT-TDDFT (kick strength $\kappa=0.001$~\AA{}$^{-1}$) and from linear-response TDDFT (LR-TDDFT) using the Casida scheme implemented in \textsc{Octopus} with the Perdew-Zunger~\cite{perd-zung81pr} LDA functional and with Gaussian16 using the LDA functional. For 1T, the second excitation is taken for the LDA (\textsc{Octopus}) results, since the first dark one is a numerical artifact. The TDDFT results are compared against experimental absorption data at room temperature in solution (acetonitrile for 1T and dioxane for the other molecules) taken from Ref.~\citenum{beck+96jpch}.}
\label{tab:benchmark_summary}
\begin{tabular}{cccccc}
\hline \hline
System & RT-TDDFT & \multicolumn{2}{c}{LR-TDDFT}    & Exp. (sol.) \\
\cmidrule(lr){3-4}
& & LDA (\textsc{Octopus}) & LDA (G16)  \\
\hline
1T & 5.5 & 5.59 & 5.89   & 5.37 \\
2T & 3.8 & 3.84 & 3.87   & 4.09 \\
4T & 2.5 & 2.54 & 2.96   & 3.16 \\
6T & 2.0 & 2.03 & 2.03   & 2.84 \\
\hline \hline
\end{tabular}
\end{table}

\clearpage

\bibliographystyle{unsrtnat}
\bibliography{references}